\documentclass[a4paper,12pt]{article}
\usepackage[utf8]{inputenc}

\renewcommand{\baselinestretch}{1.2}
\usepackage[english]{babel}
\usepackage{amsmath}
\usepackage{amsfonts}
\usepackage{amssymb}
\usepackage{graphicx}
\usepackage{bm}

\usepackage{float}
\usepackage{caption}
\usepackage{subfig}

\usepackage{booktabs}
\usepackage{comment}

\usepackage{amsthm}
\usepackage{mathtools}
\usepackage{cancel}

\theoremstyle{definition} 

\newtheorem*{defn*}{Definizione}

\theoremstyle{plain}

\newtheorem*{prop*}{Theorem}
\newtheorem*{cor*}{Corollario}
\newtheorem*{lemma*}{Lemma}

\theoremstyle{remark}

\usepackage{yfonts}
\usepackage{braket}
\usepackage{tensor}

\DeclareMathOperator{\tr}{tr}

\usepackage{cite}

\numberwithin{equation}{section}

\newcommand{\be}{\begin{equation}} \newcommand{\ee}{\end{equation}}
\newcommand{\bea}{\begin{equation} \begin{aligned}} \newcommand{\eea}{\end{aligned} \end{equation}}

\allowdisplaybreaks

\usepackage{customjheppub}

\title{The large\,-$N$ limit of 4d superconformal indices for general BPS charges}
\author[a,b]{Edoardo Colombo}
\affiliation[a]{Dipartimento di Fisica, Universit\`a di Milano-Bicocca, I-20126 Milano, Italy}
\affiliation[b]{INFN, sezione di Milano-Bicocca, I-20126 Milano, Italy}
\emailAdd{e.colombo100@campus.unimib.it}
\abstract{
We study the superconformal index of $\mathcal N=1$ quiver theories at large\,-$N$ for general values of electric charges and angular momenta,
using both the Bethe Ansatz formulation and the more recent elliptic extension method.
We are particularly interested in the case of unequal angular momenta, $J_1\ne J_2$, which has only been partially considered in the literature.
We revisit the previous computation with the Bethe Ansatz formulation with generic angular momenta 
and extend it to encompass a large class of competing exponential terms. 
In the process, we also provide a simplified derivation of the original result.
We consider the newly-developed elliptic extension method
as well; we apply it to the $J_1\ne J_2$ case, finding a good match with the Bethe Ansatz results. 
We also investigate the relation between the two different approaches,
finding in particular that for every saddle of the elliptic action there are corresponding terms in the Bethe Ansatz formula that match it at large\,-$N$.
}

\begin{document}

\maketitle

\hrule

{\renewcommand{\baselinestretch}{1} \parskip=0pt
\setcounter{tocdepth}{2}
\tableofcontents}

\vspace{0.65 cm}
\hrule


\section{Introduction}

The AdS/CFT duality offers the ideal framework to investigate
the microscopical origin of the entropy carried by 
black holes that are asymptotically\,-AdS. 
The microstates of a black hole in the bulk correspond holographically to an ensemble of states of the dual CFT on the boundary;
the black hole entropy can be determined by counting these states,
and the result should be compared with the semiclassical prediction given by the Bekenstein-Hawing formula
\cite{Bekenstein:1972tm,Bekenstein:1973ur,Bardeen:1973gs,Hawking:1974sw,Hawking:1975vcx}.
The first successful 
microstate counting of this type was obtained for 
a class of static 
dyonic BPS black holes in $\text{AdS}_4\times S^7$
\cite{Benini:2015eyy,Benini:2016hjo,Benini:2016rke}, 
and has been followed by an extensive literature.%
\footnote{
		A comprehensive list of references can be found in the review \cite{Zaffaroni:2019dhb}.}

Recently, the case of Kerr-Newman BPS black holes in $\text{AdS}_5$ 
has attracted a lot of interest.
There are known BPS black hole solutions for type IIB supergravity in $\text{AdS}_5\times S^5$ 
which depend on two angular momenta $J_{1,2}$
and three electric charges $Q_{1,2,3}$, with a non-linear constraint among them
\cite{Gutowski:2004ez,Gutowski:2004yv,Chong:2005da,Chong:2005hr,Kunduri:2006ek}.
In the weak-gravity\,/\,large\,-$N$ limit
it should be possible to reproduce the entropy of these black holes by counting
the 1/16 BPS states of the dual boundary theory, that is $\mathcal N=4\,$ Super Yang-Mills on $S^3\times\mathbb R$.
These states are counted, with a $(-1)^F$ sign, by the superconformal index \cite{Romelsberger:2005eg,Kinney:2005ej}.
However, early attempts to compute the superconformal index \cite{Kinney:2005ej,Nakayama:2005mf,Gadde:2010en,Eager:2012hx}
did not reproduce the expected $\mathcal O(N^2)$ growth,
leading to the belief that large cancellations between fermions and bosons caused by the $(-1)^F$ sign
made this approach non-viable.
Recently, 
a solution to this puzzle has been found:
when the fugacities associated to electric charges and angular momenta are extended to complex values
 the cancellations between fermions and bosons states are obstructed
\cite{Choi:2018hmj,Benini:2018ywd}.
At leading order the resulting expression for the superconformal index 
does indeed match
the \emph{entropy function} 
of $\text{AdS}_5\times S^5$ black holes
\cite{Hosseini:2017mds},
both in the large-\,$N$ limit 
\cite{Benini:2018ywd,Cabo-Bizet:2019eaf,ArabiArdehali:2019orz,Benini:2020gjh,Copetti:2020dil,Aharony:2021zkr} 
and in 
the limit of large conserved charges (i.e.\ the Cardy limit)
\cite{Choi:2018hmj,Honda:2019cio,ArabiArdehali:2019tdm,ArabiArdehali:2019orz,GonzalezLezcano:2020yeb,Goldstein:2020yvj,
	Amariti:2020jyx,Cassani:2021fyv,ArabiArdehali:2021nsx,Jejjala:2021hlt}.
The entropy function is the Legendre transform of the black hole entropy;
it can also be written as the complexified on-shell action of the Euclidean black hole geometry \cite{Cabo-Bizet:2018ehj,Cassani:2019mms}.
These results for the computation of the superconformal index have also been extended to the case of more general $\mathcal N=1$ 
gauge theories
\cite{GonzalezLezcano:2019nca,Lanir:2019abx,Cabo-Bizet:2020nkr,Benini:2020gjh}
\cite{Kim:2019yrz,Cabo-Bizet:2019osg,Amariti:2019mgp,GonzalezLezcano:2020yeb,Amariti:2021ubd,Cassani:2021fyv,ArabiArdehali:2021nsx}.

There are 
primarily two 
distinct methods 
that have been used
to compute the superconformal index of $\mathcal N=1$ quiver theories at large\,-$N$.%
\footnote{There is also the approach of \cite{Copetti:2020dil}, which considered a truncated matrix model for the index and 
		showed that higher order corrections are numerically small.}
The first 
one makes use of the \emph{Bethe Ansatz formula} 
\cite{Closset:2017bse,Benini:2018mlo},
which recasts the standard integral representation of the index \cite{Kinney:2005ej,Romelsberger:2005eg,Dolan:2008qi} 
as a sum over the solutions of a set of transcendental equations, the Bethe Ansatz equations (BAE).
The Bethe Ansatz formula simplifies considerably in the particular case of equal angular momenta; 
for this reason
the computations of \cite{Benini:2018ywd,ArabiArdehali:2019orz,Aharony:2021zkr}
for $\mathcal N=4$ Super Yang-Mills and \cite{GonzalezLezcano:2019nca,Lanir:2019abx}
for more generic quiver theories were
restricted to $J_1=J_2$.
The $J_1\ne J_2$ case was finally addressed in \cite{Benini:2020gjh}: 
a particular contribution to the Bethe Ansatz formula for the index
has been shown to reproduce the entropy of Kerr-Newman BPS black holes 
with arbitrary charges.
However, a notable limitation of \cite{Benini:2020gjh} is that we only computed a single 
exponentially growing term out of the many competing ones that contribute to the Bethe Ansatz formula.

The other approach to the large\,-$N$ computation of the  superconformal index
is the \emph{elliptic extension} method \cite{Cabo-Bizet:2019eaf,Cabo-Bizet:2020nkr,Cabo-Bizet:2020ewf}.
It
consists of a saddle point analysis
of the matrix integral representation of the index, 
with the peculiarity that the 
integrand is not extended analytically outside its contour of integration; 
instead, it is extended to a doubly periodic function.
The action of the matrix integral is thus well-defined on a torus,
and a large class of saddle point solutions can be found by taking advantage of its periodicity properties.
This method was pioneered 
in \cite{Cabo-Bizet:2019eaf} for $\mathcal N=4$ Super Yang-Mills
and later generalized to other quiver gauge theories in \cite{Cabo-Bizet:2020nkr}; 
furthermore, a reformulation of this approach that 
that is exact even at finite values of $N$ has been developed in \cite{Cabo-Bizet:2020ewf}.
So far this 
type of saddle point analysis has been 
employed 
only for the case of equal angular momenta; 
the reason behind this technical restriction is that the modulus of the torus is taken to be equal the chemical potential
of the angular momentum $J\equiv J_1=J_2$.

The primary motivation behind the work presented in this paper is to better understand
the large\,-$N$ behavior of the superconformal index for general values of BPS charges, especially in the case of unequal angular momenta, $J_1\ne J_2$.
We consider both approaches, the elliptic extension method and the Bethe Ansatz formalism, in order to provide an estimate of the large\,-$N$
limit of the index.
We also investigate the relation between the two methods,
focusing in particular on 
what the saddles of the elliptic action correspond to in the Bethe Ansatz formalism.

First, we extend the saddle point analysis of \cite{Cabo-Bizet:2019eaf,Cabo-Bizet:2020nkr}
to the $J_1\ne J_2$ case.
We achieve this 
by employing the same trick as \cite{Benini:2018mlo}:
we can assume without loss of generality that the angular chemical potentials are integer multiples of the same quantity, 
that is $\tau=a\omega$ and $\sigma=b\omega$, so that we can take advantage of the 
properties of the elliptic gamma functions \cite{Felder:1999}
to rewrite the 
action as a function that is well-defined on a torus of modulus $ab\omega$.
We find that the class of solutions to the saddle point equations described in \cite{Cabo-Bizet:2020nkr} 
can be easily generalized to the $\tau\ne\sigma$ case, 
and we compute their effective action.

We then consider the Bethe Ansatz approach to the large\,-$N$ computation of the index,
proceeding as following.
\begin{itemize}
\item
We revisit the computation of \cite{Benini:2020gjh}, 
which focused only on a single contribution to the Bethe Ansatz formula, and extend it to encompass
a large class of competing exponential terms, finding a good match with the effective action of the elliptic saddles.
Our large\,-$N$ estimate of the superconformal index is thus verified in both formalisms.
\item
We provide a simplified derivation of the same large\,-$N$ result of \cite{Benini:2020gjh}.
The most laborious step in the computation of \cite{Benini:2020gjh} is proving that
a particular simplification does not affect the large\,-$N$ leading order of the index.
We show how this step can be avoided altogether, provided that $N$ and $ab$ are coprime.
\item
We study the relation between the saddles 
of the elliptic extension method
and the  solutions to the Bethe Ansatz equations (BAE),
with the intent to shed light on 
the connection between the two different approaches.
We find that in the $J_1=J_2$ case every elliptic saddle corresponds exactly to a BAE solution; 
however this is no longer true when $J_1\ne J_2$, since the elliptic action and the Bethe Ansatz equations have different periodicities. 
Nonetheless, we show that 
matching elliptic saddles with holonomy configurations that contribute to the Bethe Ansatz formula is always possible,
as long as the role of the auxiliary integer variables $m_i$ present in the Bethe Ansatz formalism is taken into consideration.
This matching is not always exact:
sometimes 
the 
two differ by $\mathcal O(1/N)$ corrections, 
which can be shown to produce a negligible effect at leading order.
\end{itemize}

This paper is organized as follows.
In section \ref{The superconformal index of quiver theories} 
we introduce the integral representation of the superconformal index of $\mathcal N=1$ quiver theories
and we define the elliptic extension of the integrand.
In section \ref{Large-N saddle points and the effective action} we describe the saddles of the elliptic action
and compute their effective action.
In section \ref{BAsection} we switch to the Bethe Ansatz formalism: 
in subsection \ref{BAE solutions and saddle-points of the elliptic action} we 
study the relation between solutions of the Bethe Ansatz equations and saddles of the elliptic action,
while in subsection \ref{large N dominant contribution revisited} we evaluate the large\,-$N$ limit of 
the contributions to the Bethe Ansatz formula that correspond to holonomy distributions that match the saddles;
lastly, in subsection \ref{Relation with previous work} we elaborate on the relation between our results and the ones of \cite{Benini:2020gjh}.
In section \ref{Summary and discussion} we provide a summary of our results and discuss some open questions. 

\section{The superconformal index of quiver theories}
\label{The superconformal index of quiver theories}

We are interested in computing the large\,-$N$ limit of the superconformal index of a
broad class of four dimensional $\mathcal N=1$ quiver gauge theories. 
We will focus on theories 
whose gauge group can be written as the direct sum of SU($N$) subgroups,
and with matter fields that transform in either the adjoint or the bifundamental representation.
The exact field content of these theories can be summarized in the quiver diagram,
a directed graph with $|G|$ nodes and and $n_\chi$ arrows (oriented edges) between them, according to the following rules:
\begin{itemize}
	\item	Each node of the quiver denotes a SU($N$) subgroup of the gauge group $G$. 
	\item An arrow between two distinct nodes denotes a chiral multiplet in the bifundamental
		representation of the 
		two SU($N$) groups associated to the respective nodes.
	\item An arrow that has both ends attached to the same node denotes a chiral multiplet in
		the adjoint representation of the respective SU($N$) subgroup.
\end{itemize}
In this section we 
briefly review some of the generalities regarding the superconformal index; 
we introduce 
the integral representation of the index and discuss how 
to extend its integrand to a function that is well-defined on a torus.

The superconformal index of a 
$\mathcal N=1$ field theory on $\mathbb R\times S^3$ 
is defined by the following trace \cite{Kinney:2005ej,Romelsberger:2005eg}:
\be
	\mathcal I(p,q,v_a)=\tr\,(-1)^F\,e^{-\beta\{\mathcal Q,\mathcal Q^\dagger\}}\,p^{J_1+\frac R2}\,q^{J_2+\frac R2}
		\,\prod_{a=1}^{\text{rk}(G_F)}\,(v_a)^{q_a}\:,
\ee
where $\mathcal Q$ is a complex supercharge, $R$ is the generator of the R-symmetry, $J_{1,2}$ are the angular momenta relative to $S^3$,
and lastly the $q_a$ are the Cartan generators of the flavor symmetry group $G_F$.
Only  the BPS states of the theory that are annihilated by $\mathcal Q$ and $\mathcal Q^\dagger$ give a non-vanishing contribution to the trace;
in particular this means that
the index 
does not depend on the value of $\beta$.

The complex numbers $p$, $q$ and $v_a$ are the fugacities associated to the angular momenta and the flavor charges respectively.
It is useful to define the chemical potentials $\tau$, $\sigma$, $\xi_a$ by
\be
	p=e^{2\pi i\,\tau},\qquad q=e^{2\pi i\,\sigma},\qquad v_a=e^{2\pi i\,\xi_a}\:.
\ee
There is a more convenient parametrization for the flavor chemical potentials $\xi_a$. 
If $I=1,\ldots,n_\chi$ is an index that runs over all the chiral multiplets of the theory, we can set $\,\xi_I\equiv\omega_I(\xi)$,
where $\omega_I$ is the weight of the representation of $G_F$ under which the $I$-th chiral multiplet transforms.
The new chemical potentials $\xi_I$ are not linearly independent,
they satisfy the following constraint for every superpotential term $W$ in the Lagrangian:
\be
\label{sum_xi}
	\sum_{I\in W}\xi_I=0\:. 
\ee
This is a consequence of the invariance of the theory under the flavor group $G_F$.

It is convenient to define yet another set of chemical potentials as
\be
	\Delta_I=\xi_I+r_I\,\frac{\tau+\sigma}2\:,
\ee
where $r_I$ is the R-charge of the $I$-th chiral multiplet.
Notably, the superconformal index as a function of $\tau$, $\sigma$, $\Delta_I$ is invariant under integer shifts of its arguments.%
\footnote{In general the same is not true when the index is written as a function of $\tau$, $\sigma$ and $\xi_a$.
		The reason is that the index is not a single-valued function of the fugacities $p$, $q$ and $v_a$, unless all the R-charges of the theory are even.}
For each superpotential term $W$ in the Lagrangian we have that
\be
\label{sumDelta}
	\sum_{I\in W}\Delta_I=\tau+\sigma+n_W\:,
\ee
where $n_W\in\mathbb Z$ can be chosen arbitrarily, considering that the $\Delta_I$ are only defined up to integers.
This constraint follows from (\ref{sum_xi}) and the fact that each superpotential term must have R-charge 2:
\be
	\sum_{I\in W}r_I=2\:.
\ee

The superconformal index has an integral representation
\cite{Kinney:2005ej,Romelsberger:2005eg,Dolan:2008qi}, which 
for a quiver theory with 
SU($N$) groups and matter in the adjoint and bifundamental representations 
takes the following form:
\be
\label{index}
	\mathcal{I}	=\kappa\int[D\underline u\,]\:
	\prod_{\alpha=1}^{|G|}\prod_{i\ne j=1}^N\Gamma_e\Big(u_{ij}^{\alpha}+\tau+\sigma\,;\tau,\sigma\Big)\cdot
	\prod_{\alpha,\beta=1}^{|G|}\prod_{I_{\alpha\beta}}\prod_{i,j=1}^N\Gamma_e\Big(u_{ij}^{\alpha\beta}+\Delta_I\:;\tau,\sigma\Big)\:.
\ee
This formula is only valid when the chemical potentials are inside the domain
\be
\label{overall_domain}
	\text{Im}(\tau+\sigma)>\text{Im}\,\Delta_I>0\:,\qquad\text{Im}\,\tau>0\:,\qquad\text{Im}\,\sigma>0\:.
\ee
We will only consider values of the chemical potentials within this domain throughout this paper.

The integration variables in 
formula (\ref{index}) are the \emph{holonomies} $u_i^\alpha$, which parametrize the Cartan subalgebra of the gauge group;
the index $\alpha$ runs over the $|G|$ nodes of the quiver, while the index $i$ runs from 1 to $N$.
For brevity we use the notation 
$u_{ij}^\alpha\,\equiv\,u_i^\alpha-u_j^\alpha\,$ and $\,u_{ij}^{\alpha\beta}\,\equiv\,u_i^\alpha-u_j^\beta$.
Each holonomy is integrated over the interval $[\,0,1)\,$; the integration measure is given by
\be
	[D\underline u\,]=\frac1{(N!)^{|G|}}\,\prod_{\alpha=1}^{|G|}\delta\bigg(\sum_{j=1}^Nu_j^\alpha\bigg)\prod_{i=1}^Ndu_i^\alpha\:.
\ee
The delta functions ensure that the 
the matrix $\,\text{diag}\,(u^\alpha_1,\ldots,u^\alpha_N)$ belongs in the Cartan subalgebra of SU($N$) for every value of $\alpha$.

As usual $I=1,\ldots,n_\chi$ is an index that runs over the chiral multiplets of the theory.
We used the symbol $\,\prod_{I_{\alpha\beta}}$ to denote the product over a subset of the possible values of $I$,
specifically the subset that includes only the multiplets that in the quiver diagram are
represented by an arrow that goes from the node $\alpha$ to the node $\beta$.

The special function $\Gamma_e$ that appears in the integrand of (\ref{index}) is the elliptic gamma function \cite{Felder:1999}, it is defined by
\be
	\Gamma_e\Big(z\,;\tau,\sigma\Big)=\prod_{j,k=0}^\infty\frac{1-e^{2\pi i\,\big((j+1)\tau+(k+1)\sigma-z\;\!\big)}}
		{1-e^{2\pi i\,\big(j\tau+k\sigma+z\big)}}\,\:.
\ee
This infinite product is convergent for 
$\,\tau,\,\sigma\in\mathbb H\,$, where $\,\mathbb H=\{z\in\mathbb C\,\:|\:\,\text{Im}\,z>0\}$.

Lastly, the prefactor $\kappa$ that appears in front of (\ref{index}) is given by
\be
\label{kappa}
	\kappa=\left[\:\prod_{k=1}^\infty\Big(1-e^{2\pi ik\tau}\Big)\Big(1-e^{2\pi ik\sigma}\Big)\right]^{|G|(N-1)}
		\prod_{\alpha=1}^{|G|}\prod_{I_{\alpha\alpha}}\Gamma_e\Big(\Delta_I\:;\tau,\sigma\Big)^{-1}\:.
\ee
Since $\log\kappa=\mathcal O(N)$, in the large\,-$N$ limit this term gives a subleading contribution and can be neglected.

In the integral representation of the index, the contour of integration for the holonomies $u^\alpha_i$ lies exclusively on the real axis.
The integrand of (\ref{index}) can be extended analytically to the rest of the complex plane, 
since it is a product of elliptic gamma functions, which are are meromorphic.
However it is possible to consider different extensions to the complex plane;
one of the key ideas behind the saddle-point approach of \cite{Cabo-Bizet:2019eaf,Cabo-Bizet:2020nkr} for the large\,-$N$ limit of the index
is to forgo the analytic extension of the integrand in favor of a doubly periodic one. 
Focusing exclusively on the $\tau=\sigma$ case, the authors of \cite{Cabo-Bizet:2019eaf,Cabo-Bizet:2020nkr} 
rewrote the integral representation of the index in terms of the function $Q_{c,d}(z\,;\tau)$, which 
is a doubly periodic
function in $z$ with periodicities $\,1\,,\:\tau\,$ that matches the elliptic gamma function on the real axis as following:
\be
\label{different_extensions}
	Q_{c,d}\big(x\,;\tau\big)=\Gamma_e\big(x+(c+1)\tau+d\,;\tau,\tau\big)^{-1}\:,\qquad\forall\:x\in\mathbb R\:.
\ee
For all $c,d\in\mathbb R$ the $Q_{c,d}$ function is defined by \cite{Cabo-Bizet:2019eaf}
\be
\label{Qcd}
	Q_{c,d}\,(z\,;\tau)=e^{\pi i\tau\left(\frac{c^3}{3}-\frac{c}{6}\right)}\frac{Q(z+c\tau+d\,;\tau)}{P(z+c\tau+d\,;\tau)^c}\,\:,
\ee
where the functions $P$ and $Q$ are defined by (\ref{P}) and (\ref{Q}) respectively \cite{article2,article,Pa_ol_2018}.
There is an ambiguity in the definition of the phase of $P$, $Q$ which will play an important role in the discussion of section \ref{Contour deformation}.

One of the goals of this paper is to extend the computation of \cite{Cabo-Bizet:2019eaf,Cabo-Bizet:2020nkr} to the case of unequal angular momenta.
We can assume without loss of generality  that the angular chemical potential $\tau$, $\sigma$ are such that $\tau/\sigma$ is a rational number.
Indeed, the set $\{(\tau,\sigma)\in\mathbb H^2\,\:|\:\,\tau/\sigma\in\mathbb Q\}+\mathbb Z^2$ is dense in $\mathbb H^2$;
considering that the index as a function of $\tau$, $\sigma$, $\Delta_I$ is invariant under integer shifts and it is continuous,
the value of the index for generic angular chemical potentials can be inferred from the $\tau/\sigma\in\mathbb Q$ case \cite{Benini:2018mlo}.
It is then useful to define $\omega\in\mathbb H$ and integers $a$, $b$ such that
\be
	\tau=a\omega\,,\qquad\sigma=b\omega\,,\qquad\text{gcd}(a,b)=1\:.
\ee
We can now take advantage of the following gamma function identity \cite{Felder:1999}:
\be
	\Gamma_e\big(z\,;a\omega,b\omega\big)=\prod_{r=0}^{a-1}\prod_{s=0}^{b-1}\Gamma_e\big(z+(as+br)\omega\,;ab\omega,ab\omega\big)\,,
\ee
which follows from (\ref{Gamma_prod1}) and it allows us 
to write an analogue of (\ref{different_extensions}) valid for $\tau\ne\sigma$:
\be
	\Gamma_e\big(x+(c+1)ab\omega+d\,;a\omega,b\omega\big)^{-1}=
		\,\prod_{r=0}^{a-1}\prod_{s=0}^{b-1}Q_{\tfrac ra+\tfrac sb\,+c\,,\:d\,}\big(x\,;ab\omega\big)\,,\qquad\forall\:x\in\mathbb R\:.
\ee
We can use this relation to rewrite the integral representation of the index (\ref{index}) 
in terms of a new integrand which is doubly periodic but not meromorphic: 
\bea
\label{indexQ}
	\mathcal{I}=\kappa&\int[D\underline u\,]\prod_{r=0}^{a-1}\prod_{s=0}^{b-1}
		\Bigg[\prod_{\alpha=1}^{|G|}\prod_{i\ne j=1}^NQ_{\tfrac{r+1}{a}+\tfrac{s+1}{b}-1\,,\,0}\Big(u_{ij}^\alpha\,;ab\omega\Big)\cdot\\
	&\qquad\qquad\qquad\quad\!\prod_{\alpha,\beta=1}^{|G|}\prod_{I_{\alpha\beta}}\prod_{i,j=1}^N
		Q_{\tfrac{r}{a}+\tfrac{s}{b}+(\Delta_I)_2-1\,,\:(\Delta_I)_1}\Big(u_{ij}^{\alpha\beta}\,;ab\omega\Big)\Bigg]^{-1}\:,
\eea
where 
$(\Delta_I)_{1,2}\,$ are defined by 
\be
	\Delta_I\,\equiv\,(\Delta_I)_1+ab\omega(\Delta_I)_2\:,\qquad(\Delta_I)_{1,2}\in\mathbb{R}\:.
\ee
This integral representation  will be the starting point of the saddle point analysis of section \ref{Large-N saddle points and the effective action}.

\section{Large\,-$N$ saddle points and the effective action}
\label{Large-N saddle points and the effective action}

In this section we compute the large-\,$N$ limit of quiver theories for general angular momenta by   
following the same saddle-point approach as \cite{Cabo-Bizet:2019eaf,Cabo-Bizet:2020nkr}.
First, we write the matrix model (\ref{indexQ}) as
\be
	\mathcal{I}=\int[D\underline u\,]\exp\Big(-S(\underline u)\Big)\,,
\ee
where the action $S(\underline u)$ takes the following form:
\be
\label{action1}
	S(\underline u)=S_0+\sum_{\alpha=1}^{|G|}\sum_{i\ne j=1}^NV\big(u_{ij}^\alpha\;\!,\tau+\sigma\big)+
		\sum_{\alpha,\beta=1}^{|G|}\sum_{I_{\alpha\beta}}\sum_{i,j=1}^NV\big(u_{ij}^{\alpha\beta}\:\!,\Delta_I\big)\:.
\ee
Here $S_0$ is a constant that does not depend on the holonomies $\underline u$ and whose value is subleading at large\,-$N$,
while the function $V$ is defined as following:
\be
\label{action2}
	V\big(z\:\!,\Delta\big)=\sum_{r=0}^{a-1}\sum_{s=0}^{b-1}\,\log Q_{\tfrac{r}{a}+\tfrac{s}{b}+(\Delta)_2-1\,,\:(\Delta)_1}\big(z\;\!;ab\omega\big)
\ee
Since $Q_{c,d}(z;ab\omega)$ is doubly-periodic in the variable $z$ with periodicities 1 and $ab\omega$, so is the function $V$.

The saddle point equations are obtained by varying the quantity
\be
	S(\underline u,\underline{\bar u})-
		\sum_{\alpha=1}^{|G|}\sum_{i=1}^N\Big(\lambda^\alpha u_i^\alpha+\widetilde\lambda^\alpha\overline{u_i^\alpha}\Big)
\ee
with respect to the holonomies $\{u_i^\alpha\}$ and their complex conjugates $\{\overline{u_i^\alpha}\}$.
The quantities $\lambda^\alpha$ and $\widetilde\lambda^\alpha$ are Lagrange multipliers required to enforce the SU($N$) constraint.
We have denoted the action (\ref{action1}) as $S(\underline u,\underline{\bar u})$ to stress the fact that it is not meromorphic and thus
$\partial_{\,\overline{u_i^\alpha}\,}S\ne0$.
Varying with respect to $u_i^\alpha$ leads to the following equation:
\be
\label{saddle_eq}
	\sum_{j=1}^N\bigg(\partial V(u_{ij}^\alpha\;\!,\tau+\sigma)-\partial V(u_{ji}^\alpha\;\!,\tau+\sigma)+
		\sum_{\beta=1}^{|G|}\sum_{I_{\alpha\beta}}\partial V\big(u_{ij}^{\alpha\beta}\:\!,\Delta_I\big)-
		\sum_{\gamma=1}^{|G|}\sum_{I_{\gamma\alpha}}\partial V\big(u_{ji}^{\gamma\alpha}\:\!,\Delta_I\big)\bigg)=\lambda^\alpha.
\ee
Here $\partial V$ is a shorthand for $\partial_zV(z,\bar z,\Delta)$.
A similar equation with $\bar\partial V$ and $\widetilde\lambda^\alpha$ replacing $\partial V$ and $\lambda^\alpha$
is obtained when we vary with respect to $\overline{u_i^\alpha}$.

When $\,a=b=1\,$ equation (\ref{saddle_eq}) and its analogue for $\bar\partial V$ match the saddle point equations discussed in \cite{Cabo-Bizet:2020nkr}.
A large class of solutions for the $\,a=b=1\,$ case has been found in \cite{Cabo-Bizet:2019eaf,Cabo-Bizet:2020nkr} using only the periodicity properties of $V$.
When $ab\ne1$ the expression for $V$ becomes more complicated, but it still remains a doubly periodic function and thus
the solutions known for the $\,a=b=1\,$ case can be easily generalized;
we will now briefly review them.

Because of the periodicities of $V$ the solutions to equation (\ref{saddle_eq})
live in the torus 
$E_T\equiv\mathbb C/(\mathbb Z+T\,\mathbb Z)$, where $T\equiv ab\omega$.
The solutions that we consider are 
such that $u_i^\alpha=u_i^\beta\equiv u_i$ for all $\alpha,\beta$;
the advantage of this ansatz is that equation (\ref{saddle_eq}) can now be solved simply by searching for configurations $\{u_i\}_{i=1}^N$
such that the sum
\be
	\sum_{j=1}^N\partial V\big(u_j-u_i\:\!,\Delta\big)
\ee
does not depend on the value of the index $i$.
This can be achieved by taking
$\{u_i\}_{i=1}^N=\mathcal U+\bar u$, where $\mathcal U$ is a finite subgroup of the torus $E_T$ and $\bar u$ is some constant
($\bar u$ vanishes when we take the difference $u_j-u_i$).
Indeed, for any $u_i\in\mathcal U$ we have that
$\{u-u_i\}_{u\,\in\,\mathcal U\,}$ and $\,\mathcal U$ are the same set, and thus
the following sum does not actually depend on the value of $u_i$:
\be
	\sum_{u\in\mathcal U}\partial V(u-u_i\;\!,\Delta)=\sum_{u\in\mathcal U}\partial V(u\;\!,\Delta)\,,
\ee
Thus equation (\ref{saddle_eq}) is solved by the choice $\{u_i\}_{i=1}^N=\mathcal U+\bar u$,
and the same is true for the analogue equation for $\bar\partial V$.
In particular this means that these solutions to the saddle point equations can be classified by
homomorphisms of abelian groups of order $N$ into the torus $E_T$ \cite{Cabo-Bizet:2020nkr}.

Any abelian group $G$ of order $N$ is isomorphic to a product of cyclic groups:
\be
	G\,\cong\,(\mathbb Z\,/\,N_1\,\mathbb Z)\times\ldots\times(\mathbb Z\,/\,N_\ell\,\mathbb Z)\,,
\ee
where $N_1\ldots N_\ell=N$. Furthermore, we can assume without loss of generality that each $N_i$ is a divisor of $N_{i-1}$
\footnote{This is due to the fact that $(\mathbb Z\,/\,m\,\mathbb Z)\times(\mathbb Z\,/\,n\,\mathbb Z)\,\cong\,(\mathbb Z\,/\,nm\,\mathbb Z)\,$
		if $\,\text{gcd}(m,n)=1$, hence there are multiple factorizations $\,N=N_1\times\ldots\times N_\ell\,$ that, up
		to isomorphisms, define the same abelian group $G$, and it is always possible to find one that satisfies $N_i\mid N_{i-1}\:\:\forall\,i$
		\cite{Cabo-Bizet:2020nkr}.},
which we write compactly as $N_i\mid N_{i-1}$. 
The most general homomorphism of the cyclic group of order $N$ into the torus 
can be written as
\be
	i\,\mapsto\,\frac iN\big(m\,T+n\big)\,,\qquad i\in\mathbb Z\,/\,N\,\mathbb Z\,,
\ee
for some $m,n\in\mathbb Z$.
Hence, 
the most general saddle point configuration that 
corresponds to a finite group homomorphism in the torus 
takes the following form:
\be
\label{saddles}
	u_{i_1\ldots i_\ell}^\alpha=\frac{i_1}{N_1}\big(m_1\,T+n_1\big)+\ldots+\frac{i_\ell}{N_\ell}\big(m_\ell\,T+n_\ell\big)+\bar u\,,
\ee
where $N_1\ldots N_\ell=N$ and $N_i\mid N_{i-1}$. 
The value for the constant $\bar u$ is chosen so that the SU($N$) constraint is satisfied:
\be
	\sum_{i=1}^Nu_i^\alpha=0\:.
\ee
Since (\ref{action1}) only depends on differences between holonomies $\bar u$ ultimately cancels out in all the relevant equations.
From now on we will omit $\bar u$ completely.

We note that different choices of integers $\{N_i,m_i,n_i\}_{i=1}^\ell$ may lead to equivalent solutions,
that is solutions that match under the periodicities of the torus $E_T$ or permutations of the index $i$ of $u_i^\alpha$.
As an example, (\ref{saddles}) is invariant under $u_i^\alpha\mapsto-u_i^\alpha$,
or equivalently $\{m_i,n_i\}_{i=1}^\ell\mapsto\{-m_i,-n_i\}_{i=1}^\ell$.
For this reason we can assume without loss of generality that $m_1\geq0$.

\subsection{Contour deformation}
\label{Contour deformation}

It is not sufficient for the saddles (\ref{saddles}) to be stationary points of the action (\ref{action1})
for them to contribute to the integral representation of the index (\ref{indexQ});
it is necessary for the contour of integration to pass through the saddle point as well.
There is a problem: the integrand of (\ref{indexQ}) is not meromorphic, 
and thus it is not possible to use the Cauchy theorem to change contour.
An alternative procedure for the deformation of the contour has been used in \cite{Cabo-Bizet:2019eaf,Cabo-Bizet:2020nkr} 
for the analysis of the $\tau=\sigma$ case; in this section we will 
show that it can be adapted to the $\tau\ne\sigma$ case as well.

In both integral representations of the superconformal index, (\ref{index}) and (\ref{indexQ}),
each holonomy variable $u_i^\alpha$ is integrated over the interval $[\,0,1)$.
In the case of (\ref{index}) the integrand is meromorphic and we are free to deform the contour as long as we don't cross any poles;
however the saddles (\ref{saddles}) are only stationary points of the integral representation with a doubly periodic integrand (\ref{indexQ}).
The key insight of \cite{Cabo-Bizet:2019eaf,Cabo-Bizet:2020nkr} is that the integrands of (\ref{index}) and (\ref{indexQ})
are equal when evaluated on any given saddle, as long as the phase of the $Q_{c,d}$ function is chosen appropriately.
The idea is to deform the contour of integration of the meromorphic integrand to one that passes thought the saddle point, and then
show that the meromorphic integrand can be substituted with the doubly periodic one up to subleading corrections.

In order to show that the argument of \cite{Cabo-Bizet:2019eaf,Cabo-Bizet:2020nkr} can be adapted the $\tau\ne\sigma$ case
we only need to check that the integrand of (\ref{index}), which is a product of elliptic gamma functions $\Gamma_e$,
and the integrand of (\ref{indexQ}), which depends on the $Q_{c,d}$ function and the choice of its phase,
match when the holonomies $u^\alpha_i$ take (\ref{saddles}) as their value.

When $z$ is real the functions $Q_{c,d}(z;\tau)$ and $\Gamma_e(z+(c+1)\tau+d\,;\tau,\tau)^{-1}$ match exactly;
otherwise their relation is given by the following formula \cite{Cabo-Bizet:2019eaf},
obtained by substituting (\ref{Q}) in (\ref{Qcd}):
\be
\label{Q_Gamma_comparison}
	Q_{c,d}(z;\tau)=e^{2\pi i\,\alpha_Q(z+c\tau+d)}\,											
		e^{-2\pi i\tau A_c(z_2)}\,\frac{P(z+c\tau+d\,;\tau)^{z_2}}{\Gamma_e(z+(c+1)\tau+d\,;\tau,\tau)}\:.	
\ee
Here $z_{1,2}\in\mathbb R$ are defined by $z\equiv z_1+\tau\,z_2$. 
The phase of $Q_{c,d}$ depends on the particular choice for the real-valued function $\alpha_Q$. 
Apart from the constraint $\alpha_Q(x)=0\:\:\forall\:x\in\mathbb R$, $\alpha_Q$ can be chosen arbitrarily in the fundamental domain
$0\leq z_{1,2}<1\,$; its value on the rest of the complex plain is then fixed by the requirement that $Q_{c,d}(z;\tau)$
must be doubly periodic in $z$ with periods 1, $\tau$.

The rest of this subsection will be dedicated to showing that the integrands of (\ref{index}) and (\ref{indexQ})
are equal in absolute value when evaluated on any given saddle. 
It is then possible to choose $\alpha_Q$ appropriately so that the integrands match in phase as well,
and thus the contour deformation argument of \cite{Cabo-Bizet:2019eaf,Cabo-Bizet:2020nkr} can also be applied to the $\tau\ne\sigma$ case.

The function $A_c$ that appears in (\ref{Q_Gamma_comparison}) denotes the following cubic polynomial:
\be
	A_c(x)=\tfrac16x^3+\tfrac12cx^2+\tfrac12c^2x-\tfrac1{12}x\:.
\ee
We can show that the total contribution of $A_c$ to the integrand of (\ref{indexQ}) 
vanishes when evaluated at the saddle points, that is
\be
	\sum_{r,s}\sum_{i,j}\bigg[|G|\,A_{\frac{r+1}{a}+\frac{s+1}{b}-1}\big((u_{ij})_2\big)+
		\sum_IA_{\frac{r}{a}+\frac{s}{b}+(\Delta_I)_2-1}\big((u_{ij})_2\big)\bigg]\bigg|_{u_i\text{ as in }(\ref{saddles})}=0\:.
\ee
First, we note that the odd powers of $x$ in $A_c(x)$ vanish when we sum over $i,j$ since $(u_{ij})_2$ changes sign when $i$ and $j$ are exchanged.
This leaves only the quadratic term in $x$, which is proportional to $c$; when we sum over $r,s$ and all the multiplet contributions the $c$-terms vanish:
\bea
\label{trRnonanomalous}
	&\sum_{r=0}^{a-1}\sum_{s=0}^{b-1}\left[|G|\bigg(\frac{r+1}{a}+\frac{s+1}{b}-1\bigg)+
		\sum_I\bigg(\frac{r}{a}+\frac{s}{b}+(\Delta_I)_2-1\bigg)\right]=\\[2mm]
	&\qquad=\frac{a+b}2\left[|G|+\sum_I\bigg(\frac{2ab(\Delta_I)_2}{a+b}-1\bigg)\right]=0\:.
\eea
The term in the square bracket
in the second line 
can be shown to be vanishing by imposing the U(1)\textsubscript R\,-\,gauge\textsuperscript2 anomaly cancellation condition,
which for the quiver theories that we are considering can be written as following
\footnote{
		Condition (\ref{ABJ}) is equivalent to the statement that $\tr\widetilde R=\mathcal O(1)$ at large\,-$N$,
		for any R-symmetry $\widetilde R$. 
		For more details we refer to appendix B of \cite{Cabo-Bizet:2020nkr}.}:
\be
\label{ABJ}
	|G|+\sum_I\big(\;\!\widetilde r_I-1\big)=0\:.
\ee
This relation is valid for \emph{any} R-symmetry.
Then (\ref{trRnonanomalous}) follows from (\ref{ABJ}) if we consider the R-symmetry 
obtained by assigning the following charges to each chiral multiplet:
\be
	\widetilde r_I\equiv\frac{2ab(\Delta_I)_2}{a+b}\:.
\ee
Because of relation (\ref{sumDelta}) 
this choice of R-charges
does indeed satisfy
\be
	\sum_{I\in W}\,\widetilde r_I=
							2
\ee
for every superpotential term $W$ in the Lagrangian.

The contribution of $\log |P|$ to the integrand 
is vanishing as well:
\be
	\sum_{i,j=1}^N(u_{ij})_2\,\log\Big|P\big(u_{ij}+(c+1)+d\,;\tau\big)\Big|=0\:.
\ee
This relation can be derived from the double Fourier expansion of $\log|P|$ (\ref{P_Fourier})
and the fact that sums of the following type vanish:
\be
	\sum_{i_k,j_k=1}^{N_k}(i_k-j_k)\,e^{2\pi i\,\big(\frac{i_k-j_k}{N_k}\,(m_kn-n_km)\big)}=0\:.
\ee

Since the contribution of the $A_c$ and $\log|P|$ terms is overall zero, from (\ref{Q_Gamma_comparison}) we see that
the integrands of (\ref{index}) and (\ref{indexQ}) are equal in absolute value on the saddles, which is what we needed to show.

We conclude this subsection by mentioning that in \cite{Cabo-Bizet:2020ewf} a more rigorous framework for this type of saddle point
analysis has been presented, based on Atiyah-Bott-Beligne-Vergne equivariant integration formula
\cite{Duistermaat:1982vw,duistermaat1983addendum,Witten:1982im,Berline1983ZerosDC,Atiyah:1984px}.
The method of \cite{Cabo-Bizet:2020ewf} is also applicable at finite $N$, 
and it provides more solid evidence for the fact that the (\ref{saddles}) saddles do indeed contribute to index.

\subsection{Continuum limit}
\label{Continuum limit}

In the large\,-$N$ limit the saddles (\ref{saddles}) become uniform continuous distributions.
We can make the substitutions
\be
	u_i^\alpha\,\longmapsto\,u^\alpha(x)\:,\qquad\sum_{i=1}^N\:\longmapsto\:N\int_0^1dx
\ee
and replace the discrete action (\ref{action1}) with a large\,-$N$ effective action $S_\text{eff}[u]$,
which is a functional of the distribution $u^\alpha(x)$ and is given by
\be
\label{effective}
	S_{\text{eff}}[u]=N^2\int_0^1dx\int_0^1dy\,\Bigg[\,\sum_{\alpha=1}^{|G|}V\big(u^\alpha(x)-u^\alpha(y)\;\!,\tau+\sigma\big)
		+\sum_{\alpha,\beta=1}^{|G|}\sum_{I_{\alpha\beta}}V\big(u^\alpha(x)-u^\beta(y)\:\!,\Delta_I\big)\Bigg].
\ee
The stationary points of this action can be found by 
extremising the functional 
\be
	S_{\text{eff}}[u]\,-\,\sum_{\alpha=1}^{|G|}\int_0^1dx\Big(\lambda^\alpha u^\alpha(x)+\widetilde\lambda^\alpha\overline{u^\alpha(x)}\Big)\:,
\ee
and correspond to the continuum limit of the discrete saddles (\ref{saddles}).
The superconformal index at large\,-$N$ can then be written as a sum over these stationary points:
\be
	\mathcal I\,\sim\sum_{u\in\{\text{saddles}\}}\exp\big(-S_{\text{eff}}[u]\,\big)\,.
\ee

In order to take the continuum limit of the saddles (\ref{saddles}) we need to distinguish between a few cases.
Each saddle depends on a particular factorization of $N$, that is \,$N\equiv N_1\ldots N_\ell$\, with \,$N_i\mid N_{i-1}$\, $\forall\,i$\::
\be
	u_{i_1\ldots i_\ell}^\alpha=\frac{i_1}{N_1}\big(m_1\,T+n_1\big)+\ldots+\frac{i_\ell}{N_\ell}\big(m_\ell\,T+n_\ell\big)\:.
\ee
Hence, 
the $N\to\infty$ limit can be realized in multiple ways.

Let us consider the case of saddles with $\ell=1$ first.
After the the substitution $i_1/N_1\mapsto x$ they become
\be
\label{ell1}
	u^\alpha(x)=x\,\big(m\,T+n\big)\:.
\ee
We omitted the subscript on $m_1$ and $n_1$ as it is no longer needed.
The effective action for these saddles can be written as
\begin{align}
\label{S_eff_def}
\nonumber
	S_{\text{eff}}(m,n)=&\,N^2\int_0^1dx\,dy\,\bigg[|G|\,V\Big((x-y)(mT+n),\tau+\sigma\Big)+\sum_IV\Big((x-y)(mT+n),\Delta_I\Big)\bigg]\\
	=&\,N^2\int_0^1dx\,\bigg[|G|\,V\Big(x(mT+n),\tau+\sigma\Big)+\sum_IV\Big(x(mT+n),\Delta_I\Big)\bigg]\:.
\end{align}
The second equality follows from the fact that $m\,T+n$ is a period of $V$.
Another consequence of the periodicity of $V$ is that $S_{\text{eff}}(m,n)=S_{\text{eff}}(m/h,n/h)$, where $h\equiv\text{gcd}(m,n)$.
This is expected, considering that 
the $(m,n)$ saddle 
describes a distribution of 
holonomies 
equivalent to the one of the $(m/h,n/h)$ saddle. 
Thus for the $\ell=1$ ``string-like" saddles we can assume that $\text{gcd}(m,n)=1$ without loss of generality.
We postpone the computation of $S_\text{eff}(m,n)$ to section \ref{String-like saddles}.

We consider the $\ell=2$ saddles now. 
Let us first assume that $N_2\sim\mathcal O(1)$ at large\,-$N$.
We can make the substitution $i_1/N_1\mapsto x$ and write the $\ell=2$ saddles in the continuum limit as
\be
\label{ell2}
	u_{i_2}^\alpha(x)=x\,\big(m_1\,T+n_1\big)+\frac{i_2}{N_2}\big(m_2\,T+n_2\big)\:.
\ee
If we want to write the saddle without the extra index $i_2$ we can change variables to $x_\text{new}\equiv x/N_2+i_2/N_2$ 
so that
\be
\label{ell2rewrite}
	u^\alpha(x)=\{N_2\;\!x\}\,\big(m_1\,T+n_1\big)+\frac{\lfloor N_2\;\!x\rfloor}{N_2}\,\big(m_2\,T+n_2\big)\,,
\ee
where $\{N_2\;\!x\}\equiv N_2\;\!x-\lfloor N_2\;\!x\rfloor$.
It is straightforward to see that (\ref{ell2rewrite})
extremises the effective action (\ref{effective}) for any value of $N_2$.
For convenience we will use representation (\ref{ell2}) and keep the index $i_2$;
the effective action is then given by
\bea
\label{S_eff_def2}
	S_{\text{eff}}(m_1,n_1;m_2,n_2,N_2)=\frac{N^2}{N_2}\sum_{i_2=1}^{N_2}\int_0^1dx
		\,\bigg[|G|\,V\Big(x(m_1T+n_1)+\frac{i_2}{N_2}(m_2\,T+n_2),\tau+\sigma\Big)+\\
		+\sum_IV\Big(x(m_1T+n_1)+\frac{i_2}{N_2}(m_2\,T+n_2),\Delta_I\Big)\bigg].
\eea
Again, without loss of generality we can assume that $\text{gcd}(m_1,n_1)=\text{gcd}(m_2,n_2)=1$.
We postpone the computation of (\ref{S_eff_def2}) to section \ref{Saddles with multiple factors}.

If we take $N_2\to\infty$ 
in (\ref{ell2}) we obtain the ``surface" saddles:
\be
\label{surface}
	u^\alpha(x,y)=x\,\big(m_1\,T+n_1\big)+y\,\big(m_2\,T+n_2\big)\:.
\ee
The effective action for these saddles is the same as (\ref{S_eff_def2}), 
provided that the following substitutions are made:
\be
		\frac{i_2}{N_2}\:\longmapsto\:y\,\:,\qquad\frac1{N_2}\,\sum_{i_2=1}^N\:\longmapsto\:\int_0^1dy\:\,.
\ee
Because of the periodicity of the potential $V$, as long as $(m_1,n_1)$ and $(m_2,n_2)$ are linearly independent
the effective action of surface saddles does not depend on any of these integers:
\be
	\int_0^1dx\int_0^1dy\:V\Big(x(m_1T+n_1)+y(m_2T+n_2),\Delta\Big)=\int_0^1dx\int_0^1dy\:V\Big(x\,T+y\,,\Delta\Big)\:.
\ee
On the other hand if $(m_1,n_1)$ and $(m_2,n_2)$ are linearly dependent the saddle (\ref{surface}) is just equivalent
to one of the ``string-like" saddles (\ref{ell1}).

The saddles with $\ell\geq3$ in the continuum limit are always equivalent to one of the already discussed cases, (\ref{ell1}), (\ref{ell2}) or (\ref{surface}).
To see why, let us fist assume that $m_1\ne0$.
We can rewrite the $\ell=2$ saddle (\ref{ell2}) by shifting $x\mapsto x-(i_2/N_2)(m_2/m_1)$, obtaining the following equivalent expression:
\be
\label{ell2rewrite2}
	u_{i_2}^\alpha(x)=x\,\big(m_1\,T+n_1\big)+\frac{i_2}{N_2}\,\frac{m_1n_2-m_2n_1}{m_1}\:.
\ee
Similarly, a generic saddle with $\ell=3$ and $m_1\ne0$ after the $\,i_1/N_1\mapsto x\,$ substitution and analogue shifts
can be written as 
\be
\label{ell3}
	u_{i_2,i_3}^\alpha(x)=x\,\big(m_1\,T+n_1\big)+\frac{i_2}{N_2}\,\frac{m_1n_2-m_2n_1}{m_1}+\frac{i_3}{N_3}\,\frac{m_1n_3-m_3n_1}{m_1}\:.
\ee
Considering that $N_3\mid N_2\,$ and $\text{gcd}(m_1,n_1)=1$, it is always possible to find $(\widetilde m_2,\widetilde n_2)$ 
such that the $\ell=2$ saddle with $m_1,n_1,\widetilde m_2,\widetilde n_2,N_2$ is equivalent to (\ref{ell3}).
The $m_1=0$ case is similar: the saddle
\be
	u_{i_2,i_3}^\alpha(x)=x+\frac{i_2}{N_2}\big(m_2\,T+n_2\big)+\frac{i_3}{N_3}\big(m_3\,T+n_3\big)
\ee
can be rewritten as
\be
	u_{i_2,i_3}^\alpha(x)=x+\frac{i_2}{N_2}\,m_2\,T+\frac{i_3}{N_3}\,m_3\,T
\ee
by shifting $x\mapsto x-(i_2/N_2)n_2-(i_3/N_3)n_3\,$;
it is then always possible to find $\widetilde N_2$ such that the saddle is equivalent to
\be
	u^\alpha_{\,\widetilde\imath_2}(x)=x+\frac{\widetilde\imath_2}{\widetilde N_2}\,T\:.
\ee
In conclusion, there is no need to consider saddles with $\ell\geq3$ in the continuum limit.
This argument does not hold at finite $N$ however; we will discuss the saddles (\ref{saddles}) at finite $N$ in more detail in section
\ref{BAE solutions and saddle-points of the elliptic action}.

\subsection{String-like saddles}
\label{String-like saddles}

In this section we focus of the saddles $u^\alpha(x)=x(mT+n)$ and compute their effective action $S_\text{eff}(m,n)$.
Without loss of generality we can assume that $\text{gcd}(m,n)=1$ and $m\geq0$.
Given (\ref{action2}) and (\ref{S_eff_def}), 
the effective action of these saddles 
takes the following form:
\bea
	S_{\text{eff}}(m,n)=N^2\int_0^1dx\sum_{r=0}^{a-1}\sum_{s=0}^{b-1}\Bigg[&
		\,|G|\log Q_{\tfrac{r+1}{a}+\tfrac{s+1}{b}-1\,,\,0}\Big(x(mab\omega+n)\,;ab\omega\Big)+\\
		&+\sum_I\log Q_{\tfrac{r}{a}+\tfrac{s}{b}+(\Delta_I)_2-1\,,\:(\Delta_I)_1}\Big(x(mab\omega+n)\,;ab\omega\Big)\Bigg]\:.
\eea

When $m\ne0$ the integral can be computed using formula (\ref{master_fmla}), which we can write as
\be
\label{operative_integral}
	\int_0^1dx\log Q_{c,d}\Big(x(m\tau+n)\,;\tau\Big)=-\frac{\pi i}6c\tau+\frac{\pi i}3\,\frac{B_3\big([m(c\tau+d)]'_{m\tau+n}\big)}{m(m\tau+n)^2}
		+\Big(\text{purely imaginary}\Big)\,,
\ee
where the function $[\:\cdot\:]'_\tau$ is defined as follows:
\be
\label{primedbracket}
	\big[x+y\tau\big]'_\tau=
		\begin{cases}
			x-\lfloor x\rfloor+y\tau&\text{for }x\in\mathbb R\smallsetminus\mathbb Z,\,y\in\mathbb R\\
			\text{either }\,y\tau\,\text{ or }\,y\tau+1\quad&\text{for }x\in\mathbb Z,\,y\in\mathbb R
		\end{cases}\:\:.
\ee
There is an ambiguity in the definition of $[z]'_\tau$ when $z\in\mathbb Z+\tau\,\mathbb R$;
however, because of property (\ref{Bernoullitranslation}) of the Bernoulli polynomials and the fact that $B_n(0)=B_n(1)$, one can see that
equation (\ref{operative_integral}) is unaffected by this ambiguity.
The purely imaginary terms left out from equation (\ref{operative_integral})
do not actually contribute to the large\,-$N$ leading order of the effective action, considering that $S_\text{eff}$ is defined up to multiples of $2\pi i$.

Using (\ref{operative_integral}) we 
find the contribution of a single multiplet to the effective action:
\begin{align}
\label{multiplet_contribution_intermediate}
	&N^2\sum_{r=0}^{a-1}\sum_{s=0}^{b-1}
		\int_0^1dx\:\log Q_{\tfrac{r}{a}+\tfrac{s}{b}+(\Delta)_2-1\,,\:(\Delta)_1}\Big(x(mab\omega+n)\,;ab\omega\Big)=\\[3mm]\nonumber
	&\qquad=-\frac{\pi i}6\,abN^2\Big(ab\omega(\Delta)_2-\frac{\tau+\sigma}2\Big)
		+\pi iN^2\sum_{r=0}^{a-1}\sum_{s=0}^{b-1}\frac{B_3\big([m\Delta+m\omega(as+br-ab)]'_{mab\omega+n}\big)}{3m(mab\omega+n)^2},
\end{align}
where $\,\Delta\equiv\tau+\sigma\,$ for vector multiplets and $\Delta\equiv\Delta_I$ for the $I$-th chiral multiplet.
When we sum over all multiplet contributions the first term in the second line of (\ref{multiplet_contribution_intermediate})
gives an overall null contribution because of anomaly cancellation relations;
it is indeed the same (\ref{trRnonanomalous}) term that we discussed in section \ref{Contour deformation}, up to a proportionality constant.
The effective action for $(m,n)$ saddles with $m\ne0$ can thus be written as
\be
\label{1factor_result}
	S_{\text{eff}}(m,n)=\pi iN^2\bigg(|G|\,\Psi_{m,n}(\tau+\sigma)+\sum_I\Psi_{m,n}(\Delta_I)\bigg)\,,
\ee
where $\Psi_{m,n}(\Delta)$ denotes the following quantity:
\be
\label{multiplet_contribution}
	\Psi_{m,n}(\Delta)=\sum_{r=0}^{a-1}\sum_{s=0}^{b-1}\frac{B_3\big([m\Delta+m\omega(as+br-ab)]'_{mab\omega+n}\big)}{3m(mab\omega+n)^2}\:.
\ee

As a simple check, 
we notice that (\ref{multiplet_contribution}) is invariant under $(m,n)\mapsto(-m,-n)$, as expected.
Using that $\,[-z]'_\tau=1-[z]'_\tau\,$ and property (\ref{Breflect}) of the Bernoulli polynomials,
we can see that under $(m,n)\mapsto(-m,-n)$ the numerator of the summand in (\ref{multiplet_contribution}) changes sign;
since the denominator changes sign as well, (\ref{multiplet_contribution}) is indeed invariant.

When $a=b=1$ we have $\tau=\sigma=\omega$ and the effective action (\ref{1factor_result}) matches the analogous result obtained in
\cite{Cabo-Bizet:2020nkr}. 
It is also in accord with the results 
\cite{Benini:2018ywd,GonzalezLezcano:2019nca,Lanir:2019abx,ArabiArdehali:2019orz,Aharony:2021zkr} derived from the Bethe Ansatz formula.
As for the $a\ne b$ case, 
the effective action of the $(m,n)=(1,0)$ saddle matches perfectly the contribution
to the index computed in \cite{Benini:2020gjh} using the Bethe Ansatz formalism; 
we will discuss in more detail the relation between the saddle point and the Bethe Ansatz approaches in section
\ref{BAE solutions and saddle-points of the elliptic action}. 

As already noted in \cite{Benini:2020gjh}, expression (\ref{multiplet_contribution}) can be simplified significantly when $(m,n)=(1,0)$.
Using the translation property of the Bernoulli polynomials (\ref{Bernoullitranslation}) 
it is possible to write $\Psi_{1,0}(\Delta)$ as
\begin{align}
\label{Psi10expansion}
\nonumber
	\Psi_{1,0}(\Delta)=&\,\frac1{3(ab\omega)^2}\sum_{r=0}^{a-1}\sum_{s=0}^{b-1}B_3\Big([\Delta]'_\omega+\omega(as+br-ab)\Big)=\\[1mm]
	=&\,\frac13\sum_{r=0}^{a-1}\sum_{s=0}^{b-1}\sum_{k=0}^3{3\choose k}(ab\omega)^{k-2}\left(\frac{r}{a}+\frac{s}{b}+\frac{a+b}{2ab}-1\right)^k
																	B_{3-k}\Big([\Delta]'_\omega-\frac{\tau+\sigma}2\Big).
\end{align}
The sum over $r$ and $s$ can now be easily computed by means of a simple trick;
we consider the following power series
\begin{align}
\nonumber
	\sum_{k=0}^\infty\sum_{r=0}^{a-1}\sum_{s=0}^{b-1}\left(\frac{r}{a}+\frac{s}{b}+\frac{a+b}{2ab}-1\right)^k\frac{(2t)^k}{k!}&=\,
			\sum_{r=0}^{a-1}\sum_{s=0}^{b-1}\,e^{\,2t\left(\tfrac{r}{a}+\tfrac{s}{b}+\tfrac{a+b}{2ab}-1\right)}=
				\frac{\sinh^2t}{\sinh\tfrac{t}{a}\sinh\tfrac{t}{b}}=\\[3mm]
	&=ab+\frac{t^2}{6}\left(2ab-\frac{a}{b}-\frac{b}{a}\right)+O(t^4)\,.
\end{align}
which gives us immediately the relations that we need:
\be
\label{trickyfinitesum}
	\sum_{r=0}^{a-1}\sum_{s=0}^{b-1}\left(\frac{r}{a}+\frac{s}{b}+\frac{a+b}{2ab}-1\right)^k=
		\begin{cases}
			ab&\quad\text{for }\:k=0\\
			0&\quad\text{for }\:k=1\\
			\frac{1}{12}\left(2ab-\frac{a}{b}-\frac{b}{a}\right)&\quad\text{for }\:k=2\\
			0&\quad\text{for }\:k=3
		\end{cases}
\ee
Substituting (\ref{trickyfinitesum}) in (\ref{Psi10expansion}) we get
\be
\label{Psi10rewritten}
	\Psi_{1,0}(\Delta)=\frac1{3\tau\sigma}\,B_3\Big([\Delta]'_\omega-\frac{\tau+\sigma}2\Big)+
		\frac{1}{12}\Big(2ab-\frac{a}{b}-\frac{b}{a}\,\Big)B_1\Big([\Delta]'_\omega-\frac{\tau+\sigma}2\Big).
\ee
When we sum over all the multiplets, the total contribution to the effective action $S_{\text{eff}}(1,0)$ coming from the $B_1$ terms
is purely imaginary and at leading $N^2$ order can be neglected.
Indeed the $\omega$\,-\,dependent part of the $B_1$ term gives a total contribution proportional to the term in the second line of (\ref{trRnonanomalous}).
Therefore, we can equivalently define the function $\Psi_{1,0}(\Delta)$ as
\be
\label{Psi10result}
	\Psi_{1,0}(\Delta)\,\equiv\,\frac1{3\tau\sigma}\,B_3\Big([\Delta]'_\omega-\frac{\tau+\sigma}2\Big)\,.
\ee
The disappearance of the term proportional to $2ab$ 
is not surprising 
considering that the index is ultimately a continuous function of $\tau=a\omega$ and $\sigma=b\omega$.

\subsubsection*{The $\bm{(m,n)=(0,1)}$ saddle}

So far we have assumed $m\ne0\,$; let us now discuss the $m=0$ case. 
The requirement $\text{gcd}(m,n)=1$ only leaves $n=\pm1$ as possible choices, and they are equivalent; 
hence, there is only one saddle with $m\ne0$. 
As we will now show, 
the effective action of this saddle is zero at the leading $N^2$ order, 
which is coherent with the results obtained in \cite{Cabo-Bizet:2019eaf,Cabo-Bizet:2020nkr,Benini:2018ywd} for the $\tau=\sigma$ case.

For the $(m,n)=(0,1)$ saddle the $\{u_i^\alpha\}$ are all real and thus 
the doubly periodic function $Q_{c,d}$ simply coincides with the analytic elliptic gamma, and thus the action
$S(\underline u)$ given by (\ref{action1}) and (\ref{action2}) is just minus the logarithm of the integrand of (\ref{index}).
We find it easier in this case to work with the elliptic gamma functions directly rather than the $Q_{c,d}$.

First, let us look at the contribution to the effective action of the $(m,n)=(0,1)$ saddle coming from a chiral multiplet.
Using the property (\ref{Gamma_prod2}) and the definition (\ref{Gamma_def}) of the elliptic gamma function we can write it as
\begin{align}
\label{(0,1)chiral}
\nonumber
	&-\sum_{i,j=1}^N\log\Gamma_e\Big(\Delta_I+\frac{i-j}N\,;a\omega,b\omega\Big)=-N\log\Gamma_e\Big(N\Delta_I\,;Na\omega,Nb\omega\Big)=\\
	&\qquad=N\sum_{j,k=0}^\infty\Bigg[\log\bigg(1-e^{2\pi i\,N\big(ja\omega+kb\omega+\Delta_I\big)}\bigg)-
		\log\bigg(1-e^{2\pi i\,N\big((j+1)a\omega+(k+1)b\omega-\Delta_I\big)}\bigg)\Bigg].
\end{align}
If either $\big((j+1)a\omega+(k+1)b\omega-\Delta_I\big)$ or $\big(\:ja\omega+kb\omega+\Delta_I\big)$
had a negative imaginary part the respective logarithm term would be $\mathcal O(N)$ and we would get nonzero contributions at the $N^2$ order.
However in the domain (\ref{overall_domain}) the imaginary part of these terms is always positive and at large\,-$N$ all the logarithms
are exponentially suppressed.
Hence, the chiral multiplet contribution is null at the $N^2$ order.

The contribution to the effective action coming from the vector multiplets is subleading as well.
We can write it as
\be
\label{(0,1)vector}
	-|G|\sum_{i\ne j=1}^N\log\Gamma_e\Big(a\omega+b\omega+\frac{i-j}N\,;a\omega,b\omega\Big)=
		-N|G|\sum_{\ell=1}^{N-1}\log\Gamma_e\Big(a\omega+b\omega+\frac \ell N\,;a\omega,b\omega\Big)\:.
\ee
This term is of order $\mathcal O(N\log N)$. Indeed, if we substitute the definition (\ref{Gamma_def}) of the elliptic gamma function in the following product
\begin{align}
	&\prod_{\ell=1}^{N-1}\Gamma_e\Big(a\omega+b\omega+\frac iN\,;a\omega,b\omega\Big)=\\[1mm]\nonumber
	&\qquad=\prod_{\ell=1}^{N-1}\Bigg[\Big(1-e^{-2\pi i\frac\ell N}\Big)\Big(1-e^{2\pi i\left(a\omega-\frac\ell N\right)}\Big)
			\Big(1-e^{2\pi i\left(b\omega-\frac\ell N\right)}\Big)\prod_{j,k=1}^\infty\frac{1-e^{2\pi i\left(a\omega+b\omega-\frac\ell N\right)}}
				{1-e^{2\pi i\left(a\omega+b\omega+\frac\ell N\right)}}\Bigg]\:,
\end{align}
then we can use a slight modification of identity (\ref{product_trick}),
\be
	\prod_{\ell=1}^{N-1}\Big(1-e^{-2\pi i\frac\ell N}\;\!z\Big)=\frac{1-z^N}{1-z}=1+z+\ldots+z^{N-1}\:,
\ee
to conclude that
\be
	\prod_{\ell=1}^{N-1}\Gamma_e\Big(a\omega+b\omega+\frac iN\,;a\omega,b\omega\Big)
		=N\:\:\frac{1-e^{2\pi iNa\omega}}{1-e^{2\pi ia\omega}}\:\:\frac{1-e^{2\pi iNb\omega}}{1-e^{2\pi ib\omega}}=\mathcal O(N)\:,
\ee
and thus (\ref{(0,1)vector}) does not contribute to the leading $N^2$ order either. 

\subsection{General saddles} 
\label{Saddles with multiple factors}

In section \ref{String-like saddles} we considered the particular case of the $u^\alpha(x)=x(mT+n)$ saddles; 
we will now evaluate the effective action of the other 
saddles discussed in section \ref{Continuum limit}. 
Other than the surface saddles (\ref{surface}), in the continuum limit the only type of saddles that we still need to account for are the
``two-factor" saddles (\ref{ell2}),
whose effective action $S_{\text{eff}}(m_1,n_1;m_2,n_2,N_2)$ is given by (\ref{S_eff_def2}). 
We start from the two-factor saddles and postpone the discussion about surface saddles at the end of this section.

We will assume that $m_1\ne0\,$; without loss of generality we can take $m_1>0$ and $\text{gcd}(m_1,n_1)=\text{gcd}(m_2,n_2)=1$.
The contribution to the effective action coming from a single multiplet is given by the following expression:
\be
\label{multiplet2}
	\frac{N^2}{N_2}\sum_{i_2=1}^{N_2}\int_0^1dx\sum_{r=0}^{a-1}\sum_{s=0}^{b-1}
		\log Q_{\tfrac{r}{a}+\tfrac{s}{b}+(\Delta)_2-1\,,\:(\Delta)_1}\Big(x(m_1ab\omega+n_1)+\frac{i_2}{N_2}(m_2ab\omega+n_2)\,;ab\omega\Big)\,,
\ee
where as usual $\Delta$ is equal to $\tau+\sigma$ for vector multiplets and to $\Delta_I$ for the $I$-th chiral multiplet.
In order to compute (\ref{multiplet2}) we first generalize formula (\ref{operative_integral}) to include the
sum over the new index $i_2$.
Using (\ref{master_fmla}) and ignoring purely imaginary terms we find that
\begin{align}
\label{2factorintegral}
\nonumber
	&\frac1{N_2}\sum_{i_2=1}^{N_2}\int_0^1dx\log Q_{c,d}\Big(x(m_1\tau+n_1)+\frac{i_2}{N_2}(m_2\tau+n_2)\,;\tau\Big)=\\[1mm]\nonumber
	&\qquad=-\frac{\pi i}6\,c\tau+\frac{\pi i}3\,\frac1{N_2}\sum_{i_2=1}^{N_2}
		\frac{B_3\big(\{m_1d-n_1c+\frac{i_2}{N_2}(m_1n_2-m_2n_1)\}+c(m_1\tau+n_1)\big)}{m_1(m_1\tau+n_1)^2}=\\[2mm]
	&\qquad=-\frac{\pi i}6\,c\tau+\frac{\pi i}3\,\frac{B_3\big([m(c\tau+d)]'_{m\tau+n}\big)}{m(m\tau+n)^2}\,.
\end{align}
In the last equality we used formula (\ref{Bernoullisum}) to simplify the sum of Bernoulli polynomials and we defined the integers $m$ and $n$ as following:
\be
\label{mn_def}
	\big(m,n\big)\equiv\frac{N_2}{\text{gcd}(N_2,m_1n_2-m_2n_1)}\cdot\big(m_1,n_1\big)\,.
\ee
Given the similarity between the last line of (\ref{multiplet2}) and the right-hand side of (\ref{operative_integral}),
the rest of the computation is identical to the one in section \ref{String-like saddles}.

In conclusion the effective action for 
the (\ref{ell2}) saddles can also be expressed in terms of the $\Psi_{m,n}(\Delta)$
function (\ref{multiplet_contribution}) as
\be
\label{2factor_result}
	S_{\text{eff}}(m_1,n_1;m_2,n_2,N_2)=\pi iN^2\bigg(|G|\,\Psi_{m,n}(\tau+\sigma)+\sum_I\Psi_{m,n}(\Delta_I)\bigg).
\ee
The difference between this expression and (\ref{1factor_result}) lies in 
the definition of the integers $m,n$: for the latter they could be any pair of coprime integers, $\text{gcd}(m,n)=1$,
while in the case of the former they are given by (\ref{mn_def}) and $\text{gcd}(m,n)={N_2}/{\text{gcd}(N_2,m_1n_2-m_2n_1)}$.
If we set $N_2=1$ the two-factor saddles (\ref{ell2}) become simple string-like saddles (\ref{ell1});
in this case the integers $m,n$ in (\ref{mn_def}) simply match $m_1,n_1$, 
and expressions (\ref{1factor_result}) and (\ref{2factor_result}) are in agreement.
Furthermore,
in the particular case of $a=b=1\,$ the effective action (\ref{2factor_result}) matches the one computed in \cite{Cabo-Bizet:2020nkr}.

An explanation for the similarity between (\ref{2factor_result}) and (\ref{1factor_result})
can be found by 
recasting the saddles (\ref{ell2}) in a new form.
Starting from expression (\ref{ell2rewrite2}), we can 
make the following manipulations:
\bea
	u_{i_2}^\alpha(x)=&\,x\,\big(m_1\,T+n_1\big)+\frac{i_2}{N_2}\,\frac{m_1n_2-m_2n_1}{m_1}=\\[2mm]
		=&\,\{m_1x\}\left(T+\frac{n_1}{m_1}\right)+\lfloor m_1x\rfloor\frac{n_1}{m_1}+\frac{i_2}{N_2}\,\frac{m_1n_2-m_2n_1}{m_1}\mod T\,.
\eea
If we set $\,x_{\text{new}}\equiv\{m_1x\}\,$ and $\,j\equiv n_1\lfloor m_1x\rfloor(m/m_1)+i_2(m_1n_2-m_2n_1)/\text{gcd}(N_2,m_1n_2-m_2n_1)\mod m$,
we can 
thus rewrite the two-factor saddle as
\be
\label{continuum_equivalence}
	u_j^\alpha(x)=\frac jm+x\left(T+\frac nm\right)\,,
\ee
where $m,n$ are the same as in (\ref{mn_def}).

From result (\ref{2factor_result}) we find the following estimate for the large\,-$N$ limit of the superconformal index:
\be 
\label{mainresult}
	\log\mathcal I\,\gtrsim\,\underset{m\ne0}{\max_{m,n\,\in\,\mathbb Z}}
		\bigg[-\pi iN^2\bigg(|G|\,\Psi_{m,n}(\tau+\sigma)+\sum_I\Psi_{m,n}(\Delta_I)\bigg)\bigg]+o(N^2)\,,
\ee
where the maximum is taken with respect to the real part.
In regions of the parameter space where there is no maximum all the competing exponentially growing contributions to the index should be summed.
In this case information about the phase of each term would be necessary to accurately compute the index,
and that would require an analysis of the $o(N^2)$ terms.
Hence, estimate (\ref{mainresult}) does not apply in these regions.
The same can be said for the codimension-one surfaces where multiple contributions have have the same real component (i.e.\ Stokes lines).

In this paper we will not try to determine which contribution maximizes (\ref{mainresult}) in each region of the parameter space.
The large\,-$N$ phase structure of the index has been studied in the case of equal angular momenta in 
\cite{Benini:2018ywd,Cabo-Bizet:2019eaf,ArabiArdehali:2019orz,Lanir:2019abx,Cabo-Bizet:2020nkr}.

\subsubsection*{Surface saddles}

The last type of saddles that we still need to account for are the surface saddles (\ref{surface}).
Assuming that $(m_1,n_1)\ne(m_2,n_2)$, 
the following relation follows from formula (\ref{master_fmla}):
\be
\label{surface_integral}
	\int_0^1dx\,dy\:\log Q_{c,d}\Big(x(m_1\tau+n_1)+y(m_2\tau+n_2)\,;\tau\Big)=\pi i\tau\Big(\frac{c^3}{3}-\frac{c}{6}\,\Big)+
		\Big(\text{purely imaginary}\Big).
\ee
As expected there is no dependence on the specific value of the integers $m_1,n_1,m_2,n_2$.
This formula can also be found by taking the $N_2\to\infty$ limit of (\ref{2factorintegral}).
In particular this means that surface saddles 
correspond to the 
$m,n\sim\mathcal O(N)$
terms in estimate (\ref{mainresult}).

Using 
relation (\ref{surface_integral})
we can compute contribution to the effective action of the surface saddle coming from a single multiplet:
\begin{align}
\nonumber
\label{surface_contribution}
	&\frac{\pi i}{3}\,N^2\,ab\omega\sum_{r=0}^{a-1}\sum_{s=0}^{b-1}\Bigg[
	\bigg(\frac{r}{a}+\frac{s}{b}+(\Delta)_2-1\bigg)^3-\frac{1}{2}\bigg(\frac{r}{a}+\frac{s}{b}+(\Delta)_2-1\bigg)\Bigg]=\\[2mm]
	&\qquad=\frac{\pi i}{3}\,N^2\,a^2b^2\omega\bigg((\Delta)_2-\frac{a+b}{2ab}\bigg)^3-\frac{\pi i}{12}\,N^2\,(a^2+b^2)\,\omega
																					\bigg((\Delta)_2-\frac{a+b}{2ab}\bigg).
\end{align}
The sums over $r,s$ in the first line are calculated quickly with the help of relations (\ref{trickyfinitesum}).
When we sum over all multiplet contributions the second term in the second line of (\ref{surface_contribution}) sums to zero: 
it is proportional to (\ref{trRnonanomalous}).
If we define a set of 
trial R-charges $\,\widehat\Delta_{\text{trial}\,,I}\,$ 
as
\be
	\widehat\Delta_{\text{trial}\,,I}=\frac{2ab\omega(\Delta_I)_2}{\tau+\sigma}\,,
\ee
then the effective action of surface saddles can be expressed 
in terms of the cubic ’t Hooft anomaly for this trial R-symmetry:
\be
\label{surface_result}
	S_{\text{eff}}=\frac{\pi i}{24}\,\frac{(\tau+\sigma)^3}{\tau\sigma}\,\tr R^3(\widehat\Delta_{\text{trial}})\:,
\ee
where the trace is taken over the fermions of the theory.
When $\tau=\sigma$ this result matches the one of \cite{Cabo-Bizet:2020nkr}.

\section{The large\,-$N$ limit with the Bethe Ansatz formula}	
\label{BAsection}

In this section we will consider a different approach to the computation of the superconformal index at large\,-$N$.
Our starting point will not be the matrix model (\ref{index}), but rather the Bethe Ansatz formula \cite{Closset:2017bse,Benini:2018mlo}.
A contribution to the Bethe Ansatz formula that reproduces the entropy of black holes with unequal angular momenta was found in \cite{Benini:2020gjh};
in this section we will revisit the computation of \cite{Benini:2020gjh} and also expand it to include more contributions.
The results we will find reaffirm estimate (\ref{mainresult}),
thus providing a double check for the saddle-point analysis of section \ref{Large-N saddle points and the effective action}.

This section is organized as follows.
We begin with a brief review of the Bethe Ansatz formula in subsection \ref{The Bethe Ansatz formula}.
Then in subsection \ref{BAE solutions and saddle-points of the elliptic action} we study the relation between the holonomy distributions
that contribute to the Bethe Ansatz formula and the saddles (\ref{saddles}) found in \cite{Cabo-Bizet:2019eaf,Cabo-Bizet:2020nkr}.
If the reader is not interested in the technical details of subsection \ref{BAE solutions and saddle-points of the elliptic action}\,
it is possible to skip directly to subsection \ref{large N dominant contribution revisited},
in which we evaluate the large\,-$N$ limit of the index with the Bethe Ansatz formula.
Lastly, in subsection \ref{Relation with previous work} we elaborate on the relation between our results and the ones of \cite{Benini:2020gjh}.

\subsection{The Bethe Ansatz formula}
\label{The Bethe Ansatz formula}

As always we assume that the angular chemical potentials $\tau$ and $\sigma$ are integer multiples of the same quantity $\omega\in\mathbb H$,
that is $\tau=a\omega$, $\sigma=b\omega$.
The Bethe Anstatz formula recasts the integral representation (\ref{index}) of the index as following:
\be
\label{indexBA}
	\mathcal{I}=\frac{\kappa}{(N!)^{|G|}}\:\!
		\sum_{\hat u\in\text{BAE}}\sum_{\{m_i^\alpha\}=1}^{ab}\mathcal{Z}(\hat u-m\omega\,;\Delta,\tau,\sigma)\:H^{-1}(\hat u\,;\Delta,\omega)\:.
\ee

The function $\mathcal{Z}(u\,;\Delta,\tau,\sigma)$ that appears in (\ref{indexBA}) denotes the integrand of matrix model (\ref{index}),
or more accurately its analytic continuation to the complex plane with respect to the holonomies $\{u_i^\alpha\}$,
and it is given by
\be
\label{integrand}
	\mathcal{Z}(u\,;\Delta,\tau,\sigma)=\prod_{\alpha=1}^{|G|}\prod_{i\ne j=1}^N\Gamma_e\Big(u_{ij}^{\alpha}+\tau+\sigma\,;\tau,\sigma\Big)
		\cdot\prod_{\alpha,\beta=1}^{|G|}\prod_{I_{\alpha\beta}}\prod_{i,j=1}^N\Gamma_e\Big(u_{ij}^{\alpha\beta}+\Delta_I\:;\tau,\sigma\Big)\,.
\ee

The first out of the two sums in formula (\ref{indexBA}) runs over the set of inequivalent solutions to 
the following transcendental equations:
\be
\label{BAE}
	1=Q^\alpha_i(u\,;\Delta,\omega)\equiv
		e^{2\pi i\lambda^\alpha}\prod_{j=1}^{N}\:\frac{\prod_{\beta=1}^{|G|}\prod_{I_{\alpha\beta}}\exp\left(2\pi iu^\alpha_i\left(\frac{1}{2}-
			\frac1\omega{\Delta_I}\right)\right)\,\theta_0\big(-u_{ij}^{\alpha\beta}+\Delta_I\:;\omega\big)}
		{\prod_{\gamma=1}^{|G|}\prod_{I_{\gamma\alpha}}\exp\left(-2\pi iu^\alpha_i\left(\frac{1}{2}-\frac1\omega{\Delta_I}\right)\right)\,
			\theta_0\big(u_{ij}^{\alpha\gamma}+\Delta_I\:;\omega\big)}\,,
\ee
where $\lambda^\alpha$ is a Lagrange multiplier and the function $\theta_0$ is defined in (\ref{theta0def}). 
Equations (\ref{BAE}) are called Bethe Ansatz equations (BAE).
A notable property of the Bethe Ansatz operators $Q_i^\alpha$ is that they are 
doubly periodic with periods 1 and $\omega$ in each holonomy $u_j^\beta$.
Thus solutions to the BAE are equivalent if they 
match under the following identifications:
\be
\label{identifications}
	u_i^\alpha\,\sim\,u_i^\alpha+1\,\sim\,u_i^\alpha+\omega
\ee
Additionally solutions that differ by a Weyl group transformation are also considered equivalent.
In our case 
Weyl group transformation consist in permutations of the $N$ holonomies associated to each SU($N$) subgroup of the gauge group.

We will focus our attention on the class of solutions to the BAE found in \cite{Hong:2018viz},
often referred to as Hong-Liu solutions. 
Given any choice of three integers $\{p,\,q,\,r\}$ such that $\,p\cdot q=N\,$ and $\,0\leq r<q$,%
\footnote{Taking into account identifications (\ref{identifications}), 
		we could substitute $r$ with $r+nq$ in (\ref{HongLiu}) for any $n\in\mathbb Z$ 
		and the solution would be the same up to a redefinition of the index $j$,
		that is $j_{\text{new}}\equiv j+n\mod p$.
		For this reason the range of $r$ can be limited to $\,0\leq r<q$.}
the following configuration of complex holonomies solves the Bethe Ansatz equations:
\be
\label{HongLiu}
	u_{jk}^\alpha=\frac{j}{p}+\frac{k}{q}\left(\omega+\frac{r}{p}\right)+\bar u\,,
\ee
where $\,j=0,\ldots,p-1\,$ and $\,k=0,\ldots,q-1\,$ constitute a new parametrization of the index $i=1,\ldots,N$,
while $\bar u$ is a constant needed to satisfy the SU($N$) constraint
\be
	\sum_{i=1}^Nu_i^\alpha=0\:.
\ee
We point out that 
the Hong-Liu solutions (\ref{HongLiu}) are such that $u_{j_1k_1}^\alpha\ne u_{j_2k_2}^\alpha\mod1,\,\omega\,$ whenever $(j_1,k_1)\ne(j_2,k_2)$,
or in other words 
they are not invariant under nontrivial Weyl group transformations. 
As argued in \cite{Benini:2018mlo}, BAE solutions that do not fit this requirement %
give an overall null contribution to the superconformal index
when plugged in the Bethe Anstatz formula (\ref{indexBA}).

Other than the discrete class of solutions (\ref{HongLiu})
there is evidence in favor of the existence of 
other solutions to the BAE, either isolated or belonging to continuous families of solutions
\cite{ArabiArdehali:2019orz,Lezcano:2021qbj,Benini:2021ano}. 
In this paper we will not 
account for the contribution of these ``non-standard" solutions, we will instead focus on the standard Hong-Liu solutions exclusively.

The other sum that appears in formula (\ref{indexBA}) is a sum over a collection of integers $\{m_i^\alpha\}$. 
When $i\ne N$ the possible values that $m_i^\alpha$ can take range from 1 to $ab\,$; 
on the other hand $m_N^\alpha$ is fixed by the SU($N$) constraint:
\be
\label{m_N}
	m_N^\alpha=-\sum_{i=1}^{N-1}m_i^\alpha\:.
\ee
However in the large\,-$N$ limit we can ignore this constraint and set $m_N^\alpha$ to whatever is most convenient:
the leading order of $\log\mathcal Z(u\,;\Delta,\tau,\sigma)$
is unaffected by a change in value of a single holonomy $u_i^\alpha$,
and thus changing $m_N^\alpha$ from (\ref{m_N}) to something else entirely does not impact the computation of the index \cite{Benini:2020gjh}.

Lastly, the quantity $H(u\,;\Delta,\omega)$ that appears in formula (\ref{indexBA}) is a Jacobian given by
\be
\label{H}
	H(u\,;\Delta,\omega)=\det\left[\frac{1}{2\pi i}\:\frac{\partial\big(\log Q_1^1,\ldots,\log Q_N^1,\ldots,\log Q_1^{|G|},\ldots,\log Q_N^{|G|}\big)}
		{\partial\big(u_1^1,\ldots,u_{N-1}^1,\lambda^1,\ldots,u_1^{|G|},\ldots,u_{N-1}^{|G|},\lambda^{|G|}\big)}\right].
\ee
In this expression the holonomies 
$\{u_N^\alpha\,|\,\alpha=1,\ldots,|G|\}$ are not considered independent variables, they are instead treated like functions of the other holonomies,
$u_N^\alpha\equiv-\sum_{i=1}^{N-1}u_i^\alpha$.
The Lagrange multipliers $\lambda^\alpha$ on the other hand are regarded as independent variables.

\subsection{BAE solutions and saddle points of the elliptic action}
\label{BAE solutions and saddle-points of the elliptic action}

For a direct comparison of the saddle point analysis with the Bethe Ansatz formula it is important to understand the relation between
the saddles found in \cite{Cabo-Bizet:2019eaf,Cabo-Bizet:2020nkr} 
with the configurations that arise from the discrete solutions to the Bethe Ansatz equations;
this will be the goal of this section.
The bulk of the computation of the large\,-$N$ limit of the index will be in section \ref{large N dominant contribution revisited},
and it is possible for the reader to skip ahead.

%
In the first half of this section we will
show that the saddles given by (\ref{saddles}) can always be written in a form similar to 
the Hong-Liu solutions (\ref{HongLiu}),
namely it is possible to find integers $p$, $q$ and $r$ and a new set of indices $j=0,\ldots,p-1$ and $k=0,\ldots,q-1$ such that%
\footnote{The vice versa however does not hold: we will later provide an explicit example of a choice of integers $p$, $q$ and $r$
		such that the right-hand side of (\ref{config_equivalence}) does not correspond to any of the saddles given by (\ref{saddles}).}
\be
\label{config_equivalence}
	u^\alpha_{i_1\ldots i_\ell}\equiv\frac{i_1}{N_1}\big(m_1\,T+n_1\big)+\ldots+\frac{i_\ell}{N_\ell}\big(m_\ell\,T+n_\ell\big)\,=
		\,\frac{j}{p}+\frac{k}{q}\left(T+\frac{r}{p}\right)\mod1,\,T\,.
\ee
This expression generalizes relation (\ref{continuum_equivalence}), which is valid only in the continuum limit, to the case of finite $N$. 
The main difference between the right-hand side of (\ref{config_equivalence}) and the BAE solutions (\ref{HongLiu})
is that the saddles of the doubly periodic action have $T\equiv ab\omega$ as their period,
while the solutions to the Bethe Ansatz equations have periodicity $\omega$.
We will address this discrepancy in the second half of this section, where we will discuss the role played the vector of integers $m$ that appears in
the Bethe Ansatz formula (\ref{indexBA}).

We can ignore without loss of generality saddle point configurations that repeat values,
or in other words saddles such that $\,u_{i_1\ldots i_\ell}^\alpha=u_{j_1\ldots j_\ell}^\alpha\mod1,\,T\,$ for some $(i_1,\ldots i_\ell)\ne(j_1,\ldots,j_\ell)$.
Since the saddles given by (\ref{saddles}) can be thought as homomorphisms of finite abelian groups into the torus,
repetitions occur only if the kernel is nontrivial. 
If the kernel contains $n$ elements, then 
the image group in the torus is the same as the image group of a 
SU($N/n$) saddle point configuration with no repetitions.
Therefore 
(\ref{config_equivalence}) holds for these saddles as long as we take $p\cdot q=N/n$,
assuming (\ref{config_equivalence}) is true for saddles that don't repeat values.
Furthermore, we note that solutions to the Bethe Ansatz equations that repeat values give an overall null
contribution to the index because they are not invariant under nontrivial Weyl group transformations.
For these reasons 
we will only consider configurations without repetitions from now on.

For $\ell=1$ the relation (\ref{config_equivalence}) has already been proven in \cite{Benini:2018ywd}.
The idea is to take  $p=\text{gcd}(m_1,N_1)$, $q=N_1/p$ and defining the new indices $k=0,...,q-1$, $\hat\jmath=0,...,p-1$
so that $i_1=sk+q\:\!\hat\jmath\,\text{ mod }N_1$, where $s$ is 
a positive integer such that $s\;\!m_1/p\,\text{ mod }\,q=1$; such an 
integer must exist since $m_1/p$ and $q$ are coprime. 
Furthermore, $s$ cannot have factors in common with $q$, and thus the set
$\{sk+q\:\!\hat\jmath\:|\:k=0,...,q-1,\,\hat\jmath=0,...,p-1\}$ covers all residue classes modulo $N_1$ once.
The saddle can then be written as
\be
\label{ell=1}
	\frac{i_1}{N_1}\big(m_1\,T+n_1\big)=\big(sk+q\:\!\hat\jmath\,\big)\left(\frac{m_1/p}q\,T+\frac{n_1}{N_1}\right)=
		\frac{n_1\hat\jmath}{p}+\frac kq\left(T+\frac{n_1s}p\right)\mod1,\,T\,,
\ee
which matches the right-hand side of (\ref{config_equivalence}) for $r\equiv n_1s\text{ mod }q$, $j\equiv n_1\hat\jmath\text{ mod }p$.

We can now prove (\ref{config_equivalence}) in the general case using induction. 
Let us assume that there are positive integers $p_1$, $q_1$ and $r_1$ such that $p_1q_1=N_1\ldots N_{\ell-1}=N/N_\ell$ 
and
\be
	\frac{i_1}{N_1}\big(m_1\,T+n_1\big)+...+\frac{i_{\ell-1}}{N_{\ell-1}}\big(m_{\ell-1}\,T+n_{\ell-1}\big)\,=\,
		\frac{j_1}{p_1}+\frac{k_1}{q_1}\left(T+\frac{r_1}{p_1}\right)\mod1,\,T\,.
\ee
The left-hand side of this identity 
is missing 
the following piece:
\be
	\frac{i_\ell}{N_\ell}\big(m_\ell\,T+n_\ell\big)\,\equiv\,\frac{j_2}{p_2}+\frac{k_2}{q_2}\left(T+\frac{r_2}{p_2}\right)\mod1,\,T\,,
\ee
where  the integers $p_2$, $q_2$ and $r_2$ are determined as in the $\ell=1$ case.
In this case $p_2$, $q_2$ satisfy $p_2q_2=N_\ell\;\!$; furthermore
the condition $N_\ell\mid N_{\ell-1}$ implies that $p_2q_2\mid p_1q_1$.
The left-hand side of (\ref{config_equivalence}) can thus be written as
\be
\label{sumofBAEsolutions}
	\frac{j_1}{p_1}+\frac{k_1}{q_1}\left(T+\frac{r_1}{p_1}\right)+\frac{j_2}{p_2}+\frac{k_2}{q_2}\left(T+\frac{r_2}{p_2}\right)\mod1,\,T\,.
\ee
We will now show that this expression doesn't repeat values only when $p_1,p_2,q_1,q_2$ satisfy $\text{gcd}(p_1,p_2)=\text{gcd}(q_1,q_2)=1$.

The necessity of the condition $\text{gcd}(p_1,p_2)=1$ can be inferred just from the $j_1/p_1+j_2/p_2$ portion of (\ref{sumofBAEsolutions}),
considering that the set
\be
	\left\{\frac{j_1p_2+j_2p_1}{\text{gcd}(p_1,p_2)}\,\:\bigg|\:j_1=0,\ldots,p_1-1,\:j_2=0,\ldots,p_2-1\right\}
\ee
covers every residue class modulo $p_1p_2/\text{gcd}(p_1,p_2)$ exactly $\text{gcd}(p_1,p_2)$ times.
Therefore if $p_1$ and $p_2$ were not coprime $(j_1/p_1+j_2/p_2\mod1)$ would repeat values, and so would (\ref{sumofBAEsolutions}) for fixed $k_1$ and $k_2$.

It is easy to see that requiring $\text{gcd}(q_1,q_2)=1$ in addition to $\text{gcd}(p_1,p_2)=1$ is sufficient to ensure that (\ref{sumofBAEsolutions})
doesn't repeat values. Indeed if $q_1$ and $q_2$ were coprime $(k_1/q_1+k_2/q_2)T$ modulo $T$ wouldn't repeat and therefore
all possible combinations of $j_1,j_2,k_1,k_2$ would give rise to unique values for expression (\ref{sumofBAEsolutions}).
On the other hand it is a little trickier to show that the condition $\text{gcd}(q_1,q_2)=1$ is necessary,
as we need to take in account the terms proportional to $r_1$ and $r_2$ as well.
Since $j_1/p_1+j_2/p_2$ covers all multiples of $1/p_1p_2$ modulo 1 once, if (\ref{sumofBAEsolutions}) doesn't have repetitions modulo 1, $T$
then the same expression without $j_1/p_1+j_2/p_2$ won't have  repetitions modulo $1/p_1p_2\,$, $T$.
Each possible value of $(k_1/q_1+k_2/q_2)T$ modulo $T$ is repeated $\text{gcd}(q_1,q_2)$ times,
which means that either $\text{gcd}(q_1,q_2)=1$ or the following expression doesn't have repetitions:
\be
	\frac{k_1r_1}{p_1q_1}+\frac{k_2r_2}{p_2q_2}\mod\,\frac1{p_1p_2}\:=\:
		\frac1{p_1q_1}\left(k_1r_1+k_2r_2\:\frac{p_1q_1}{p_2q_2}\mod\frac{q_1}{p_2}\right).
\ee
Considering that both $p_1q_1/p_2q_2$ and $q_1/p_2$ are integers%
\footnote{The condition $\text{gcd}(p_1,p_2)=1$ together with $p_2q_2\mid p_1q_1$ implies that $p_2\mid q_1$.}
and that the pair $(k_1,k_2)$ can take a total of $q_1q_2$ distinct values,
the term in the parenthesis will 
take the same values multiple times
unless $p_1=q_1=1$, which cannot be possible as it would imply $N_1=\ldots =N_{\ell-1}=1$.
Therefore we must have $\text{gcd}(q_1,q_2)=1$.

Let us now show that (\ref{sumofBAEsolutions}) can be written in the same form as the right-hand side of (\ref{config_equivalence}),
assuming that $\text{gcd}(p_1,p_2)=\text{gcd}(q_1,q_2)=1$.
First we define $k\equiv k_1q_2+k_2q_1\text{ mod }q_1q_2$; since $q_1$ and $q_2$ are coprime $k$ is an index that runs from 0 to $q_1q_2-1$ once.
Let us ignore for the moment terms that are integer multiples of $1/p_1p_2$; we can write the rest as
\be
\label{kpart}
	\frac{k_1}{q_1}\left(T+\frac{r_1}{p_1}\right)+\frac{k_2}{q_2}\left(T+\frac{r_2}{p_2}\right)=\,
		\frac1{q_1q_2}\left(kT+\frac{k_1r_1p_2q_2}{p_1p_2}\right)\mod\,\frac1{p_1p_2}\,,\:T\,.
\ee
The term proportional to $r_2$ is a multiple of $1/p_1p_2$, considering that $p_2q_2\mid p_1q_1$ and $\text{gcd}(q_1,q_2)=1$
imply that $q_2\mid p_1$. 
Since $q_1$ and $q_2$ don't have factors in common it is possible to find an integer $n$ such that $r_1p_2+nq_1=0\mod q_2\:\!$,
which is going to help us rewrite (\ref{kpart}) solely in terms of $k$: 
\be
\label{k1part}
\begin{alignedat}{3}
	k_1r_1p_2q_2=&\,k_1q_2\big(r_1p_2+nq_1\big)&&\mod q_1q_2\\
	=&\,\big(k-k_2q_1\big)\big(r_1p_2+nq_1\big)&&\mod q_1q_2\\
	=&\,k\big(r_1p_2+nq_1\big)&&\mod q_1q_2\,.
\end{alignedat}
\ee
Defining $p\equiv p_1p_2$, $q\equiv q_1q_2$ and $r\equiv r_1p_2+nq_1\text{ mod }q$, equation (\ref{kpart}) 
becomes
\bea
	\frac{k_1}{q_1}\left(T+\frac{r_1}{p_1}\right)+\frac{k_2}{q_2}\left(T+\frac{r_2}{p_2}\right)=&\,\,
		\frac kq\left(T+\frac rp\right)\mod\,\frac1p\,,\:T\equiv\\[2mm]
	\equiv&\,\,\frac kq\left(T+\frac rp\right)+\frac{n_k}p\mod1\,,\:T
\eea
for some $k$-dependent integer $n_k$. At last we can define $j\equiv j_1p_2+j_2p_1+n_k\text{ mod }p$,
so that
\be
	\frac{j_1}{p_1}+\frac{k_1}{q_1}\left(T+\frac{r_1}{p_1}\right)+\frac{j_2}{p_2}+\frac{k_2}{q_2}\left(T+\frac{r_2}{p_2}\right)=
		\,\frac{j}{p}+\frac{k}{q}\left(T+\frac{r}{p}\right)\mod1,\,T\,,
\ee
which concludes the proof of (\ref{config_equivalence}).

Vice versa, we can show that there exist some choices of integers $p$, $q$ and $r$ such that the set of points 
\be
	\left\{\frac{j}{p}+\frac{k}{q}\left(T+\frac{r}{p}\right)\:\bigg|\:j=0,\ldots,p-1,\,k=0,\ldots,q-1\right\}
\ee
does not match any of the saddles given by (\ref{saddles}), modulo 1, $T$.
One way to see this is to look at the greatest common divisor of $p$, $q$ and $r$ obtained by the procedure above.

First, in the case of saddles with $\ell=1$ the steps outlined in (\ref{ell=1}) lead to values of $p$, $q$ and $r$ that don't have factors in common:
\be
	\text{gcd}\big(p,q,r\big)=\text{gcd}\big(m_1,N_1,q,n_1s\big)=1\,,
\ee
where we used that $\text{gcd}(m_1,n_1)=1=\text{gcd}(q,s)$.
For more general saddles on the other hand we find that 
\be
\label{gcd(p,q,r)}
	\text{gcd}\big(p,q,r\big)=N_2\ldots N_\ell\:.
\ee
We can show this by means of induction, 
by writing $\text{gcd}(p,q,r)$ in terms of $\text{gcd}(p_1,q_1,r_1)$.
Considering that $kr=k_1r_1p_2q_2\mod q$ for all possible values of $k$, we 
must have that $\text{gcd}(q,r)=\text{gcd}\big(q,r_1p_2q_2\big)$.
The greatest common divisor of $p$, $q$, $r$ is thus given by
\be
\label{set_pqr}
	\text{gcd}\big(p,q,r\big)=\text{gcd}\big(p,q,r_1p_2q_2\big)=p_2q_2\,\,\text{gcd}\left(\frac{p_1}{q_2},\frac{q_1}{p_2},r_1\right)=
		p_2q_2\,\,\text{gcd}\big(p_1,q_1,r_1\big)\,.
\ee
In the last step we used that $\text{gcd}(p_1,p_2)=\text{gcd}(q_1,q_2)=1$.
Since $p_2q_2=N_\ell$, by induction we find formula (\ref{gcd(p,q,r)}).

Let us consider for example the case $p=4$, $q=4$, $r=2$.
We want to try to find a saddle point configuration that matches the set (\ref{set_pqr}) for these values of $p$, $q$, $r$.
Using formula (\ref{gcd(p,q,r)}) we find that such a saddle would have $N_2\ldots N_\ell=\text{gcd}(p,q,r)=2$, which implies $\ell=2$ and $N_2=2$.
Consequently, there are only two possible values that $\text{gcd}(m_2,N_2)$ can take, either 1 or 2, 
and neither of them works: 
the former would lead to $q_1=q_2=2$, while the latter would lead to $p_1=p_2=2$, and in both cases $\text{gcd}(p_1,p_2)=\text{gcd}(q_1,q_2)=1$
is not satisfied.
Hence there is no saddle that reproduces the configuration with $p=4$, $q=4$, $r=2$.

\subsubsection*{The different periodicities} 

The most jarring difference between the saddles (\ref{saddles}) of the doubly-periodic action and the Hong-Liu solutions (\ref{HongLiu})
to the Bethe Ansatz equations lies in the different value for $T$, the modulus of the torus, which is $ab\omega$ for the former and just $\omega$
for the latter.

In the particular case of equal angular momenta we have $\tau=\sigma\equiv\omega$,
which implies $a=b=1$ and thus
this discrepancy between the known saddles and the standard BAE solutions disappears.

When $ab\ne1$ the solution to the problem comes from a key element that we haven't taken in consideration yet:
the presence of the vector of integers $m$ in the Bethe Ansatz formula (\ref{indexBA}).
Each of its entries $m_i^\alpha$ takes values that range from 1 to $ab$, 
and its shifts the corresponding holonomy inside the argument of the integrand $\mathcal Z\,$ as $\,\hat u^\alpha_i-m_i^\alpha\omega$,
where $\hat u$ is the BAE solution that we are considering.
Rather than trying to match the saddles (\ref{saddles}) with the Hong-Liu solutions directly,
it is more sensible to compare them with configurations of the type $\hat u-m\omega$.
That is, given any choice of integers $\{p,q,r\}$, we search for a BAE solution $\hat u$ and a choice of vector $m$ such that 
\be
\label{bridging_the_gap}
	\frac jp+\frac kq\left(ab\omega+\frac rp\right)=\hat u_i^\alpha-m_i^\alpha\omega+\text{constant}\mod1,\,ab\omega\,.
\ee
The constant term ultimately vanishes because the integrand (\ref{integrand}) only depends on differences between holonomies. 

We point out that 
there is a large number of valid 
$\hat u-m\omega$ configurations other than the ones that satisfy (\ref{bridging_the_gap}).
We won't try to account for all possible $(\hat u,m)$ combinations,
especially considering that the number of possible values that the vector $m$ can take is $(ab)^{|G|(N-1)}$, which grows exponentially with $N$.

In order to find a $(\hat u,m)$ combination that satisfies (\ref{bridging_the_gap}) for a given choice of $\{p,q,r\}$, we need to
search for integers $\widetilde p$, $\widetilde q$ and $\widetilde r$ that satisfy $\widetilde p\,\widetilde q=pq$
and a new set of indices $\widetilde\jmath=0,\ldots,\widetilde p-1$ and $\widetilde k=0,\ldots,\widetilde q-1$ such that
\be
\label{idealscenario}
	\frac jp+\frac kq\left(ab\omega+\frac rp\right)
		=\frac{\widetilde\jmath}{\widetilde p}+\frac{\widetilde k}{\widetilde q}\left(\omega+\frac{\widetilde r}{\widetilde p}\right)\mod 1,\,\omega\,.
\ee
Unfortunately this isn't always possible; to see why, let us set $h\equiv\text{gcd}(q,ab)$ and define a new parametrization of the index $k$
in terms of new indices $k'=0,\ldots,q/h-1$ and $k''=0,\ldots,h-1$ such that $k\equiv k'+(q/h)\;\!k''$.
The left-hand side of (\ref{idealscenario}) then becomes
\be
\label{htrick}
	\frac jp+\frac kq\left(ab\omega+\frac rp\right)
		=\frac jp+\frac{k'}q\left(ab\omega+\frac rp\right)+\frac{k''r}{hp}\mod 1,\,\omega\,.
\ee
If $r$ and $h$ are not coprime then the right hand side of (\ref{htrick}) manifestly repeats values when $k''$ varies while $j$ and $k'$ are fixed.
This means that unless $\text{gcd}(q,r,ab)=1$ it is not possible to 
match the right-hand side of (\ref{idealscenario}), since the latter never repeats values modulo 1, $\omega$ as $\widetilde\jmath$ and $\widetilde k$ vary.

Even if it is not possible to find a $(\hat u,m)$ combination that satisfies (\ref{bridging_the_gap}) when 
$q$ and $r$ are such that $\text{gcd}(q,r,ab)\ne1$,
it is always possible to find $\hat u$ and $m$ that approximate the left-hand side of (\ref{bridging_the_gap}) well enough in the large\,-$N$ limit; 
we will discuss this in more detail in appendix \ref{General N}.

Let us consider the case $\text{gcd}(q,r,ab)=1$ and show that it is indeed possible to 
obtain (\ref{idealscenario}) starting from (\ref{htrick}).
Since $ab/h$ and $q/h$ are coprime
$k'(ab/h)\text{ mod }q/h$ takes all the values from 0 to $q/h-1$ once; therefore if we set $\widetilde q\equiv q/h$ and
$\widetilde k\equiv k'(ab/h)\text{ mod }\widetilde q\,$ we can match the $\omega$-dependent portion of (\ref{idealscenario}) and (\ref{htrick}) 
as follows:
\be
\label{omegaportion}
	\frac{k'}q\,ab\omega=\frac1{q/h}\left(k'\,\frac{ab}h\right)\omega=\frac{\widetilde k}{\widetilde q}\,\omega\mod\omega\,.
\ee
Since $k'$ also appears in the $\omega$-independent term proportional to $r$, we need to rewrite this term as well in terms of the new index $\widetilde k$;
to do so, we will ignore for the moment the role of integer multiples of $1/hp$ so that we can write
\be
	\frac{k'\:\!r}{pq}=\frac{k'\:\!(r+n\:\!q/h)}{pq}\mod\,\frac1{hp}\:=
		\left(\frac1{q/h}\:k'\,\frac{ab}h\right)\frac1{hp}\left(\frac{r+n\:\!q/h}{ab/h}\right)\mod\,\frac1{hp}\,,
\ee
where $n$ is an arbitrary integer that we have introduced.
Once again we make use of the fact that $ab/h$ and $q/h$ are coprime: 
$r+n\;\!q/h$ for $n\in\mathbb Z$ covers all the residue classes modulo $ab/h$, %
which means that we can always choose $n$ such that $\widetilde r\equiv(r+n\:\!q/h)/(ab/h)$ is an integer.
Setting $\widetilde p\equiv hp$ and noticing that the other term inside parentheses is equal to $\widetilde k\;\!/\;\!\widetilde q\text{ mod }1$, we find that
\be
\label{r-term-rewrite}
	\frac{k'\:\!r}{pq}=\frac{\widetilde k\,\widetilde r}{\widetilde p\,\widetilde q}\mod\,\frac1{\widetilde p}\:\equiv\,
		\frac{\widetilde k\,\widetilde r}{\widetilde p\,\widetilde q}+\frac{n_k}{\widetilde p}\mod1
\ee
for some $k$-dependent integer $n_k$.
We notice that the way $\widetilde p$ and $\widetilde q$ have been defined is such that $\widetilde p\,\widetilde q$ is equal to $pq$, as it should be.
At last we make use of the fact that we are assuming that $\text{gcd}(q,r,ab)=1$, which is equivalent to the statement that $h$ and $r$ are coprime,
and thus 
$\,\widetilde\jmath\equiv k''r+hj+n_k\,\text{ mod }\widetilde p\,$ is a proper definition for an index that runs from 0 to $\widetilde p-1\,$ a single time;
using (\ref{r-term-rewrite}) and 
the definition of the new index $\widetilde\jmath$ we can match the $\omega$-independent portion of (\ref{idealscenario}) and (\ref{htrick}):
\be
	\frac jp+\frac{k'\:\!r}{pq}+\frac{k''r}{hp}=\frac{\widetilde\jmath}{\widetilde p}+\frac{\widetilde k\,\widetilde r}{\widetilde p\,\widetilde q}\mod 1\,.
\ee
This concludes the proof 
of the existence of integers $\{\widetilde p,\widetilde q,\widetilde r\}$ 
for which the rewrite (\ref{idealscenario}) is possible,
under the assumption that $\text{gcd}(q,r,ab)=1$.

\subsection{Evaluation of the index}
\label{large N dominant contribution revisited}

In this section we 
will evaluate
the contribution to the Bethe Ansatz formula 
(\ref{indexBA}) coming from
distributions of holonomies of the following type:
\be
\label{config}
	(u-m\omega)_{jk}^\alpha\,=\,\frac{j}{p}+\frac{k}{q}\left(ab\omega+\frac{\widehat r}{p}\right)+\text{const}\,.
\ee
As usual $p\cdot q=N\,$ and $\,0\leq\widehat r<q\,$;
we have added the hat on $\,\widehat r\,$ in order to avoid confusion with the index $r$ that appears in definition (\ref{multiplet_contribution}).
For simplicity 
when we take the large\,-$N$ limit we will keep $p$ and $\widehat r$ fixed and send $q\to\infty$.

In section \ref{BAE solutions and saddle-points of the elliptic action} we have shown that
configurations like (\ref{config}) are possible only 
when $\text{gcd}(ab,q,\widehat r\,)=1\,$.
Throughout this section we will assume that $q$ satisfies this condition;
in appendix \ref{General N} we will show how this 
restriction can be removed.

The quantity that we need to compute is the following:
\be
\label{contributionBA}
	\lim_{q\to\infty}\,\log\Big(\,\kappa\:(N!)^{-|G|}\:\mathcal{Z}(u-m\omega\,;\Delta,\tau,\sigma)\:H^{-1}(u\,;\Delta,\omega)\,\Big)\Big|_{(\ref{config})}\,.
\ee
The prefactor $\kappa$ 
is given by (\ref{kappa}); as
already mentioned in section \ref{The superconformal index of quiver theories}, it is subleading at large\,-$N$,
specifically $\,\log\kappa=\mathcal O(N)$.
The factor $(N!)^{-|G|}$ is subleading as well, it is $\mathcal O(N\log N)$ by Stirling formula.
We are left with the Jacobian $H$, given by (\ref{H}), and the integrand $\mathcal Z$, given by (\ref{integrand}).

Let us start from the Jacobian:
we can show that generically it gives a subleading contribution 
and can be neglected,
using an argument similar to the one given in \cite{Benini:2018ywd,Lanir:2019abx}.
The Jacobian $H$ is the determinant of the matrix whose elements are the partials derivatives of
the Bethe Ansatz operators $Q^\alpha_i$ defined in (\ref{BAE}); let us examine these partial derivatives.
First, the derivatives of $Q^\alpha_i$ with respect to the Lagrange multipliers are simply 
\be
	\frac{\partial\log Q^\alpha_i}{\partial \lambda^\beta}=2\pi i\,\delta_{\alpha\beta}\,.
\ee
We can ignore these terms as they are just $\mathcal O(1)$.
On the other hand, the derivatives of with respect to the holonomies are given by
\begin{align}
\label{partialQ}
	\frac{\partial\log Q^\alpha_i}{\partial u_j^\beta}=&\,-\delta_{\alpha\beta}\,\big(\delta_{ij}-\delta_{iN}\big)\sum_{k=1}^N\sum_{\gamma=1}^{|G|}
		\Bigg[\sum_{I_{\alpha\gamma}}F(-u_{ik}^{\alpha\gamma}+\Delta_I)+\sum_{I_{\gamma\alpha}}F(u_{ik}^{\alpha\gamma}+\Delta_I)\Bigg]+\\[1mm]
\nonumber
	&+\sum_{I_{\alpha\beta}}\bigg[F(-u_{ij}^{\alpha\beta}+\Delta_I)-F(-u_{iN}^{\alpha\beta}+\Delta_I)\bigg]
		+\sum_{I_{\beta\alpha}}\bigg[F(u_{ij}^{\alpha\beta}+\Delta_I)-F(u_{iN}^{\alpha\beta}+\Delta_I)\bigg]\,,
\end{align}
where $F$ is the following function:
\be
	F(u)\equiv\frac{2\pi i}\omega\,u-\pi i+\frac{\partial_u\theta_0(u;\omega)}{\theta_0(u;\omega)}\,.
\ee
This function becomes singular only at the zeros of the $\theta_0$, that is when $u\in\mathbb Z+\omega\,\mathbb Z$.
If the the chemical potentials $\Delta_I$ are such that the distribution of points $u_{ij}^{\alpha\beta}+\Delta_I$ doesn't accumulate around any of these poles
in the limit $N\to\infty$, then $F(u_{ij}^{\alpha\beta}+\Delta_I)\sim\mathcal O(1)$ and 
\be
	\frac{\partial\log Q^\alpha_i}{\partial u_j^\beta}=\delta_{\alpha\beta}\,\big(\delta_{ij}-\delta_{iN}\big)\cdot\mathcal O(N)\,+\,\mathcal O(1)\,.
\ee
In our case $u_i^\alpha$ is given by (\ref{config}), and in the $q\to\infty$ limit the poles of $F$ are 
provided that
\be
\label{Stokes}
	\Delta_I\notin\frac{\text{gcd}(ab,\widehat r)}{pab}\,\mathbb Z+(pab\omega+\widehat r)\,\mathbb R\,.
\ee
As long as 
this condition
is satisfied for all chemical potentials only the diagonal elements and the $i=N,\,\alpha=\beta$ elements are of order $\mathcal O(N)$,
while all the others are just $\mathcal O(1)$.
In particular this means that the determinant (\ref{H}) grows like $N^{(|G|N)}$, and thus $\log H=\mathcal O(N\log N)$.

Since the Jacobian is subleading as long as the chemical potentials satisfy (\ref{Stokes}),
the large\,-$N$ leading order of (\ref{contributionBA}) is determined solely by the integrand $\mathcal Z$. 
The computation boils down to the evaluation of the following quantity:
\be
\label{Phi}
	\Phi_{p\:\!,q,\:\!\widehat r}(\Delta)\equiv\sum_{j_1,j_2=0}^{p-1}\:\sum_{k_1\ne k_2=0}^{q-1}
		\log\Gamma_e\left(\Delta+\frac{j_1-j_2}p+\frac{k_1-k_2}q\left(ab\omega+\frac{\widehat r}p\right);a\omega,b\omega\right).
\ee
In terms of this function we can write $\log\mathcal Z$ as%
\footnote{ 
		If $p$ is fixed as $N\to\infty$ then the sum of all the terms 
		that have $k_1=k_2$ is subleading:
		
		$q|G|\sum_{j_1\ne j_2=0}^{p-1}\log\Gamma_e\big(a\omega+b\omega+\frac{j_1-j_2}p\,;a\omega,b\omega\big)
			+q\sum_I\sum_{j_1,j_2=0}^{p-1}\log\Gamma_e\big(\Delta+\frac{j_1-j_2}p\,;a\omega,b\omega\big)=\mathcal O(N)$
		}
\be
	\log\mathcal{Z}(u-m\omega\,;\Delta,\tau,\sigma)\Big|_{(\ref{config})}=\,
		|G|\,\Phi_{p\:\!,q,\:\!\widehat r}(\tau+\sigma)+\sum_I\Phi_{p\:\!,q,\:\!\widehat r}(\Delta_I)+\mathcal O(N)\,.
\ee
A quick comparison with (\ref{2factor_result}) tells us that in the large\,-$N$ limit we should expect $\Phi_{p\:\!,q,\:\!\widehat r}(\Delta)$
to be related to the function $\Psi_{m,n}(\Delta)$ defined in (\ref{multiplet_contribution}).
More specifically, because of (\ref{continuum_equivalence}) we expect the leading order of $\Phi_{p\:\!,q,\:\!\widehat r}(\Delta)$
to be equal to $-\pi iN^2\:\Psi_{p,\:\!\widehat r}(\Delta)$.

Using formula (\ref{Gamma_prod2}) we can take care of the sum over $j_1,j_2$:
\be
\label{Phi_intermediate}
	\Phi_{p\:\!,q,\:\!\widehat r}(\Delta)=\:
		p\sum_{k_1\ne k_2=0}^{q-1}\log\Gamma_e\left(p\Delta+\frac{k_1-k_2}q\,\big(pab\omega+\widehat r\big)\,;pa\omega,pb\omega\right).
\ee
In order to compute the $q\to\infty$ limit of this sum 
we can take advantage of the following result found in \cite{Benini:2018ywd}:
\be
\label{Old_result}
	\sum_{i\ne j=1}^N\log\Gamma_e\left(\Delta+\frac{i-j}{N}\,\omega\,;\omega,\omega\right)=
		-\pi iN^2\frac{B_3\big([\Delta-\omega\:\!]'_\omega\:\!\big)}{3\:\!\omega^2}+o(N^2)\,.
\ee
Here the subleading terms are of order $\mathcal O(N)$ when $\Delta\notin\mathbb Z+\mathbb R\cdot\omega$,
and $\mathcal O(N\log N)$ for the $\Delta=0$ case.
We can recast (\ref{Phi_intermediate}) in a form similar to (\ref{Old_result}) by making use of the following identity:
\be
	\log\Gamma_e\Big(z\:;pa\omega,pb\omega\Big)=
		\sum_{r=0}^{a-1}\sum_{s=0}^{b-1}\log\Gamma_e\Big(z+p\omega(as+br)\,;pab\omega+\widehat r,pab\omega+\widehat r\Big)\,,
\ee
which follows from (\ref{Gamma_prod1}) and the invariance of the elliptic gamma under integer shifts of any of its arguments. 
Denoting $pab\omega+\widehat r\,$ as $\,\widetilde\omega$ for convenience, (\ref{Phi_intermediate}) becomes
\be
	\Phi_{p\:\!,q,\:\!\widehat r}(\Delta)=\,p\sum_{r=0}^{a-1}\sum_{s=0}^{b-1}\,\sum_{k_1\ne k_2=0}^{q-1}
		\log\Gamma_e\left(p\Delta+p\omega(as+br)+\frac{k_1-k_2}q\:\widetilde\omega\:;\,\widetilde\omega,\,\widetilde\omega\right).
\ee
Applying formula (\ref{Old_result}), we get at last the following 
expression for the large\,-$N$ leading order of $\Phi_{p\:\!,q,\:\!\widehat r}(\Delta)\,$:
\begin{align}
\label{BAresult}
\nonumber
	\Phi_{p\:\!,q,\:\!\widehat r}(\Delta)=&\,-\frac{\pi iN^2}{3p(pab\omega+\widehat r)^2}\sum_{r=0}^{a-1}\sum_{s=0}^{b-1}
		B_3\Big(\big[p\Delta+p\omega(as+br-ab)\big]'_{pab\omega+\widehat r}\Big)+o(N^2)=\\[2mm]
	=&\,-\pi iN^2\:\Psi_{p,\:\!\widehat r\:\!}(\Delta)+o(N^2)\,,
\end{align}
where as usual the function $\Psi_{m,n}(\Delta)$ is defined by (\ref{multiplet_contribution}).

This result is consistent with the ones we obtained in sections \ref{String-like saddles} and \ref{Saddles with multiple factors} with the elliptic extension approach.
In particular, we find that the large\,-$N$ estimate for the index (\ref{mainresult}) is verified by the Bethe Ansatz formalism as well.

As already mentioned in section 
\ref{String-like saddles}, 
the $p=1$, $\,\widehat r=0\,$ case matches the contribution to the Bethe Ansatz formula computed in \cite{Benini:2020gjh},
which reproduces the entropy function of AdS\textsubscript5 black holes.
In the next subsection we will elaborate more on how result (\ref{BAresult}) and the one of \cite{Benini:2020gjh} compare.

\subsection{Relation with previous work 
			and competing exponential terms}
\label{Relation with previous work}

Let us 
discuss the relation between the 
computation 
of subsection \ref{large N dominant contribution revisited}
and the one of \cite{Benini:2020gjh},
which also estimated the large\,-$N$ leading order of the index 
with $\tau\ne\sigma$ using the Bethe Ansatz formula. 

In \cite{Benini:2020gjh} we focused exclusively on a single contribution to the index, the one coming from
the following distribution of holonomies:
\be
\label{old_config}
	(u-m\omega)_i^\alpha\,=\,\frac iN\,\omega+(i\text{ mod }ab)\,\omega+\text{const}\,.
\ee
Up to $\mathcal O(1/N)$ terms, this configuration matches the right-hand side of (\ref{config}) with $p=1$, $\widehat r=0$.
For clarity, let us assume that $N$ is a multiple of $ab\,$; 
we can then reparametrize  the index $i$ in terms of new indices $i'=0,\ldots,N/ab-1$ and $i''=0,\ldots,ab-1$ such that $i\equiv i'ab+i''$,
and rewrite (\ref{old_config}) as following:
\be
\label{old_approx}
	\frac iN\,\omega+(i\text{ mod }ab)\,\omega\,=\,\frac{i'+i''(N/ab)}{N}\,ab\omega+\frac{i''}N\,\omega\,\equiv\,
		\frac{\widetilde\imath}N\,ab\omega\,+\,\mathcal O\left(\frac1N\right).
\ee
In the second step we defined a new index $\,\widetilde\imath=0,\ldots,N-1\,$ as $\,\widetilde\imath\,\equiv\,i'+i''(N/ab)$.
As argued in appendix A of \cite{Benini:2020gjh}, these $\mathcal O(1/N)$ terms can be 
neglected in the large\,-$N$ limit, if we are only interested in the leading order. 
Accordingly,
the contribution to the index coming from the distribution of holonomies (\ref{old_config}) computed in \cite{Benini:2020gjh}
does indeed match the result that we have obtained for the $p=1$, $\widehat r=0$ case.

Proving that the $\mathcal O(1/N)$ terms in (\ref{old_approx}) do not affect the large\,-$N$ leading order
is possibly the most laborious step in the large\,-$N$ computation of \cite{Benini:2020gjh}.
In this section we have shown that it is possible to avoid this step completely by choosing a different set up for $u-m\omega$, i.\,e.\ (\ref{config}),
at least
as long as 
the assumption $\text{gcd}(ab,q,\widehat r\,)=1\,$ 
is valid.
Since the superconformal index is a continuous function of $\tau=a\omega$ and $\sigma=b\omega$,
it is natural to expect 
that the $\text{gcd}(ab,q,\widehat r\,)=1\,$ condition doesn't actually play a role in the large\,-$N$ behavior of the index; 
rather, this condition should be a byproduct of 
focusing strictly on holonomy distributions that can be written as (\ref{config}).%
\footnote{
In appendix \ref{General N} we will verify that this intuition is indeed correct:
for any possible choice of integers $p$, $q$, $\widehat r\,$ 
there are contributions to the Bethe Anstatz formula that in the large\,-$N$ limit give the same result as (\ref{BAresult}),
even when the condition $\text{gcd}(ab,q,\widehat r\,)=1\,$ is not satisfied;
the price to pay is that we will have to deal with the $\mathcal O(1/N)$ terms once again.
}

We want to stress the fact that the distributions of holonomies (\ref{old_config}) and  (\ref{config}) (with $p=1$, $\widehat r=0$) are distinct from one another,
even if in the large\,-$N$ limit they differ just by $\,\mathcal O(1/N)$ terms.
This raises a problem: the contributions to the index coming from these two distributions  are exponentially growing terms
whose logarithms match at leading $N^2$ order, and it is easy to see
that there are many other similar contributions;%
\footnote{	For example, as already argued in \cite{Benini:2020gjh} changing the value of a single holonomy does not impact the large $N$ leading order,
		and it is always possible to change the value of a single holonomy by by changing the value of one of the entries of the vector of integers $m$.
		Hence, for any contribution to the index that we have computed there are always many possible competing exponential terms.
		}
all these competing exponential terms must be summed together 
since there 
is no guarantee that one of them clearly dominates over the others.

Let us first estimate how many competing exponential terms there are. 
Any possible choice of $u-m\omega$ 
that matches $\frac iN\,ab\omega$ up to $\mathcal O(1/N)$ terms must have $u_i\equiv \frac iN\,\omega$,
since this is the only Hong-Liu solution (\ref{HongLiu}) whose holonomies are strictly proportional to $\omega$.
There are then $(ab)^{|G|(N-1)}$ possible choices for the vector $m$, at most.
If we were to assume that all the competing exponential terms interfere constructively,
we would get at most a $|G|(N-1)\log(ab)$ correction to our previous estimate for the large\,-$N$ limit,
which is subleading and thus negligible.
In other words the 
leading $N^2$ order does not receive corrections from the multiplicity of the competing exponentials.
However it would be possible, albeit very unlikely, for all these terms to interfere destructively in such a way 
that 
they cancel completely. In order to determine whether this is the case, we would need to 
calculate the exact phase of all the competing contributions, which is unfeasible.
Given that in the saddle point analysis of section \ref{Large-N saddle points and the effective action}
this problem does not occur at all, we are lead to believe that 
such a cancellation does not happen and the leading $N^2$ order is unaffected.

\section{Summary and discussion}
\label{Summary and discussion}

In this paper we have estimated the large\,-$N$ limit of the superconformal index of $\mathcal N=1$ quiver theories
with adjoint and bifundamental matter for general values of BPS charges,
using both the elliptic extension approach of \cite{Cabo-Bizet:2019eaf,Cabo-Bizet:2020nkr,Cabo-Bizet:2020ewf}
and the Bethe Ansatz formula \cite{Closset:2017bse,Benini:2018mlo}.
We have found a good accord between the two methods,
resulting in the following estimate for the index:
\be
\label{mainresult2}
	\log\,\mathcal I\,\big(\{\Delta_I\};\tau,\sigma\big)\,\gtrsim\,\underset{m\ne0}{\max_{m,n\,\in\,\mathbb Z}}
		\bigg[-\pi iN^2\bigg(|G|\,\Psi_{m,n}(\tau+\sigma)+\sum_I\Psi_{m,n}(\Delta_I)\bigg)\bigg]+o(N^2)\,,
\ee
where the function $\Psi_{m,n}(\Delta)$ is defined by
\be
	\Psi_{m,n}(\Delta)\equiv\sum_{r=0}^{a-1}\sum_{s=0}^{b-1}\frac{B_3\big([m\Delta+m\omega(as+br-ab)]'_{mab\omega+n}\big)}{3m(mab\omega+n)^2}
		\qquad(\tau\equiv a\omega\,,\:\sigma\equiv b\omega)
\ee
and the parentheses $[\,\cdot\,]'_T$ are such that $[\,x+y\,T\,]'_T=x-\lfloor x\rfloor+y\,T$ for real $x,\,y$.

Our results extend the saddle point analysis of \cite{Cabo-Bizet:2019eaf,Cabo-Bizet:2020nkr} to the case of unequal angular momenta ($\tau\ne\sigma$).
They also extend the computation of \cite{Benini:2020gjh} to include multiple competing exponentially-growing contributions to the Bethe Ansatz formula;
the single contribution computed in \cite{Benini:2020gjh} corresponds to the $m=1$, $n=0$ term in (\ref{mainresult2}).

In section \ref{BAE solutions and saddle-points of the elliptic action}
we have shown that the saddles of the elliptic action found in \cite{Cabo-Bizet:2019eaf,Cabo-Bizet:2020nkr} can always be written in the form
\be
	\frac{j}{p}+\frac{k}{q}\left(T+\frac{r}{p}\right)+\text{const.}
\ee
for some integers $p$, $q$, $r$, with $T\equiv ab\omega$.
This is the same form that the 
standard Hong-Liu solutions \cite{Hong:2018viz} to the Bethe Ansatz equations (BAE) take, 
with the only difference being that the latter are defined on a torus with a modulus $T\equiv\omega$.
When $a=b=1$ this means that
each saddle has a matching BAE solution, 
and thus a corresponding term in the Bethe Ansatz equations;
however for general $a,b$ the different values of $T$ cause a mismatch between saddles and BAE solutions.
In this paper we have shown how the two different pictures can be reconciled: we have to consider that
each contribution to the Bethe Ansatz formula is labeled not only by the BAE solution $u$ but also by the choice of value for the auxiliary integer 
parameters $\{m_i\}$ that shift the BAE solution as $u_i\mapsto u_i-m_i\,\omega\,$.
We have found that for each saddle of the elliptic action there is a $(u,\{m_i\})$ combination that
matches it, either exactly or up to $\mathcal O(1/N)$ corrections that are negligible at large\,-$N$.

There are still some open questions 
concerning the matching between the two approaches.
Most notably, 
the number of $(u,\{m_i\})$ combinations that label each contribution to the Bethe Ansatz formula is exponentially bigger than the number of known
saddles of the elliptic action.
In this paper we have computed only the contribution of the $(u,\{m_i\})$ combinations that match a saddle,
but there are many other contributions that are unaccounted for.
It is not feasible to try to evaluate all of them, given their exponentially large number: the integers $\{m_i\}$ can take $(ab)^{|G|(N-1)}$
different values.
Furthermore, the formulas that we have used in section \ref{large N dominant contribution revisited} would not apply in general.
Nonetheless, trying to understand what role do all these terms play remains an interesting question.
The simplest possible answer would be that only the $(u,\{m_i\})$ combinations that match one of the elliptic saddles up to negligible corrections
give a contribution that at large\,-$N$ dominates in some region of the space of parameters;
further work is however still needed to test the correctness of such a conjecture.

In this paper we have not analyzed which contribution maximizes (\ref{mainresult2}) in each region of the parameter space.
A detailed study of the phase structure of the index at large\,-$N$ for general values of BPS charges 
is a possible direction for future research.
In our analysis we focus exclusively on the $\mathcal O(N^2)$ leading order;
an interesting generalization would be to compute some lower order corrections.

\acknowledgments

We thank Alberto Zaffaroni for useful discussions and comments.
We thank the organizers and participants of the SCGP seminar series on “Supersymmetric
black holes, holography and microstate counting” for interesting talks and discussions.
EC is partially supported by the INFN and the MIUR-PRIN contract 2017CC72MK003.

\appendix

\section{The elliptic gamma and related functions}
\label{functions}


\subsubsection*{The elliptic gamma function}

The elliptic gamma function \cite{Felder:1999} is 
defined by the following infinite product:
\be
\label{Gamma_def}
	\Gamma_e\Big(z\,;\tau,\sigma\Big)=\prod_{j,k=0}^\infty\frac{1-e^{2\pi i\,\big((j+1)\tau+(k+1)\sigma-z\;\!\big)}}
		{1-e^{2\pi i\,\big(j\tau+k\sigma+z\big)}}\,,
\ee
which is convergent as long as $\,\text{Im}\,\tau>0\,$ and $\,\text{Im}\,\sigma>0$.
It is meromorphic in $z$, with poles in $z\in\mathbb Z+\tau\,\mathbb Z_{\leq0}+\sigma\,\mathbb Z_{\leq0}\,$ 
and zeros in $z\in\mathbb Z+\tau\,\mathbb Z_{\geq1}+\sigma\,\mathbb Z_{\geq1}$.
It is manifestly invariant under integer shifts in $z,\tau,\sigma$ and symmetric under the exchange of $\tau$ and $\sigma$.

The elliptic gamma satisfies the following inversion relation:
\be
\label{Gamma_inversion}
	\Gamma_e\Big(z\,;\tau,\sigma\Big)=\Gamma_e\Big(\tau+\sigma-z\,;\tau,\sigma\Big)^{-1}\,,
\ee
and the following product formulas:
\begin{align}
\label{Gamma_prod1}
	\prod_{k=0}^{n-1}\Gamma_e\Big(z+\frac{k}{n}\:\!\tau\,;\tau,\sigma\Big)=&\,\Gamma_e\Big(z\:;\frac{\tau}{n},\sigma\Big)\,,\\
\label{Gamma_prod2}
	\prod_{j=0}^{m-1}\Gamma_e\Big(z+\frac{j}{m}\,;\tau,\sigma\Big)=&\,\Gamma_e\Big(mz\,;m\tau,m\sigma\Big)\,.
\end{align}
Relations (\ref{Gamma_inversion}) and (\ref{Gamma_prod1}) follow directly from definition (\ref{Gamma_def});
as for relation (\ref{Gamma_prod2}), it is 
a consequence of the following polynomial identity:
\be
\label{product_trick}
	\prod_{j=0}^{m-1}\Big(1-e^{2\pi i(j/m)}z\Big)=1-z^m\:.
\ee

\subsubsection*{The $\theta_0$ function}

The q-theta function $\theta_0$ is defined as follows:
\be
\label{theta0def}
	\theta_0(z\,;\tau)=\prod_{k=0}^\infty\left(1-e^{2\pi i(z+k\tau)}\right)\left(1-e^{2\pi i(-z+(k+1)\tau)}\right)\,.
\ee
It is analytic in $z$ and its zeros are in $z\in\mathbb Z+\tau\,\mathbb Z$.
It is related to the elliptic gamma function by the following shift identity:
\be
	\Gamma_e(z+\tau\,;\tau,\sigma)=\theta_0(z\,;\sigma)\,\Gamma_e(z\,;\tau,\sigma)\:.
\ee
On the other hand the shift identity for the $\theta_0$ itself is the following:
\be
\label{theta_shift}
	\theta_0(z+\tau\,;\sigma)=-\,e^{-2\pi i\,z}\:\theta_0(z\,;\sigma)\:.
\ee

\subsubsection*{Bernoulli polynomials}

The Bernoulli polynomials are defined by the following generating function: 
\be
	\frac{te^{xt}}{e^t-1}=\sum_{n=0}^\infty B_n(x)\,\frac{t^n}{n!}\,\:.
\ee
They satisfy the following relations:
\begin{align}
\label{Bshift}
	B_n(x+1)=&\,B_n(x)+nx^{n-1}\,,\\		
\label{Breflect}
	B_n(1-x)=&\,(-1)^nB_n(x)\,.			
\end{align}
As a consequence of (\ref{Breflect}) the Bernoulli polynomials are either even or odd when expressed in the variable $2x-1$;
in particular the first few polynomials can be written as
\bea
\label{B1and3}
	B_1(x)=&\:\frac12\,\big(2x-1\big)\\[2mm]
	B_2(x)=&\:\frac14\,\big(2x-1\big)^2-\frac1{12}\\[2mm]
	B_3(x)=&\:\frac18\,\big(2x-1\big)^3-\frac18\,\big(2x-1\big)\:.
\eea
Some useful identities include the translation property,
\be
\label{Bernoullitranslation}
	B_n(x+y)=\sum_{k=0}^n{n\choose k}B_k(x)\,y^{n-k}\,,
\ee
and the multiplication formula,
\be
\label{Bernoullimult}
	\sum_{j=0}^{m-1}B_n\left(z+\frac jm\right)=m^{1-n}\,B_n\Big(mz\Big)\,.
\ee
The multiplication formula can also be written as follows:
\be
\label{Bernoullisum}
	\sum_{j=0}^{m-1}B_n\left(z+\left\{x+\frac jm\right\}\right)=m^{1-n}\,B_n\Big(mz+\{mx\}\Big)\,,
\ee
where the brackets $\{\cdot\}$ denote the fractional part, which is defined by $x\equiv\lfloor x\rfloor+\{x\}$.
Relation (\ref{Bernoullisum}) follows directly from (\ref{Bernoullimult}) 
if we consider that
\be
	m\cdot\min_j\left\{x+\frac jm\right\}=\{mx\}\:.
\ee

\subsubsection*{The $P$, $Q$ 
						functions}

Given $z,\,\tau\in\mathbb C$, $\text{Im}\,\tau>0$, throughout the rest of this appendix we will denote with $z_1$ and $z_2$
the real numbers such that $z\equiv z_1+\tau\,z_2$.

The function $P$ is defined by
\cite{article2}
\be
\label{P}
	P(z\,;\tau)=e^{2\pi i\,\alpha_P(z)}\:e^{\pi i\tau B_2(z_2)}\:\theta_0(z\,;\tau)\,,
\ee
where 
$\alpha_P$ can be any real-valued function that satisfies the following two constraints: 
$\alpha_P$ should vanish on the real axis, and it must be chosen so that $P(z\,;\tau)$
is invariant under translations by the lattice $\mathbb Z+\tau\,\mathbb Z$ in $z$.
The second requirement can always be fulfilled because $|P|$ is manifestly invariant under integer shifts,
and it is also invariant under shifts in $\tau$, once (\ref{theta_shift}) is taken into consideration.
This 
can also be seen from the second  Kronecker limit formula \cite{article2}:
\be
\label{P_Fourier}
	\log|P(z;\tau)|=-\lim_{s\to1}\frac{(\text{Im}\tau)^s}{2\pi}\underset{(m,n)\ne(0,0)}{\sum_{m,n\in\mathbb Z}}
		\frac{e^{2\pi i(nz_2-mz_1)}}{|m\tau+n|^{2s}}\:.
\ee

Similarly, the function $Q$ is defined by 
\cite{article,Pa_ol_2018} 
\be
\label{Q}
	Q(z\,;\tau)=e^{2\pi i\,\alpha_Q(z)}\:e^{2\pi i\left(\frac13B_3(z_2)-\frac12z_2B_2(z_2)\right)}\:\frac{P(z\,;\tau)^{z_2}}{\Gamma_e(z+\tau\,;\tau,\tau)}\,,
\ee
where $\alpha_Q$ is a real-valued function chosen according to the same criteria as $\alpha_P$.
Hence $Q$ is a doubly periodic function in $z$ as well, with periods 1 and $\tau$.
Its double Fourier expansion is given by \cite{Cabo-Bizet:2019eaf}
\be
\label{Q_Fourier}
	\log Q(z\,;\tau)=-\frac1{4\pi^2}\underset{m\ne0}{\sum_{m,n\in\mathbb Z}}\frac{e^{2\pi i(nz_2-mz_1)}}{m(m\tau+n)^2}
		+\frac{2\pi i\tau}3B_3(\{z_2\})+\pi i\phi(z)\,,
\ee
where $\phi$ is a real-valued doubly periodic function related to $\alpha_Q$.

Let $m,n$ be integers such that $\text{gcd}(m,n)=1$ and $m\ne0\,$;
from (\ref{P_Fourier}) and (\ref{Q_Fourier}) it is possible to derive the following integral formulas
\cite{Cabo-Bizet:2019eaf,Cabo-Bizet:2020nkr}:
\bea
\label{fouriercoeff}
	\int_0^1dx\log P\Big(x(m\tau+n)+c\tau+d\,;\tau\Big)=&-\pi i\,\frac{B_2(\{md-nc\})}{m(m\tau+n)}+\pi i\,\varphi_P(m,n)\,,\\
	\int_0^1dx\log Q\Big(x(m\tau+n)+c\tau+d\,;\tau\Big)=&\:\frac{\pi i}3\,\frac{B_3(\{md-nc\})}{m(m\tau+n)^2}+\pi i\,\varphi_Q(m,n)\:.
\eea
Here $\phi_P$ and $\phi_Q$ are real functions whose precise value depends on the particular choice of the phases $\alpha_P$ and $\alpha_Q$.
We point out that $B_n(\{\cdot\})$ is a continuous function on the real axis because identity (\ref{Bshift}) implies that $B_n(0)=B_n(1)$.

From (\ref{fouriercoeff}) and definition (\ref{Qcd}) we can find 
a similar formula for the $Q_{c,d}$ function (\ref{Qcd}); 
making use of the identity (\ref{Bernoullitranslation}) for the Bernoulli polynomials we can write it as
\begin{align}
\label{master_fmla}
\nonumber
	&\int_0^1dx\log Q_{c,d}\Big(x(m\tau+n)+z\,;\tau\Big)=\\[2mm]
	&\qquad=-\frac{\pi i}6\,c\tau+\frac{\pi i}3\,\frac{B_3\big(\{m(d+z_1)-n(c+z_2)\}+c(m\tau+n)\big)}{m(m\tau+n)^2}\,-\\[2mm]\nonumber
	&\qquad\quad\,-\frac{\pi i}m\,c^2B_1\big(\{m(d+z_1)-n(c+z_2)\}\big)-\frac{\pi in}{3m}\,c^3-\pi i\big(\phi_Q(m,n)-c\,\phi_P(m,n)\big)\,.
\end{align}
All the terms in the last line of this equation are purely imaginary.

\section{Subleading terms in the Bethe Ansatz formula}
\label{General N}

In section \ref{large N dominant contribution revisited}
we assumed for simplicity that the integers $q$ and $\widehat r$ satisfy $\text{gcd}(ab,q,\,\widehat r\,)=1$;
we will now 
show how 
the large\,-$N$ computation of the superconformal index with the Bethe Ansatz formula can be done without this assumption.

When $\text{gcd}(ab,q,\,\widehat r\,)\ne1$ the problem is that it is not possible to find a BAE solution $u$
and a valid choice for the vector of integers $m$ such that $u-m\omega$ satisfies (\ref{config}).
The workaround is to search for $u$ and $m$ that approximate the right-hand side of (\ref{config}) instead;
to be precise we want to find a choice of integers $\{\widetilde p,\widetilde q,\widetilde r\}$,
indices $\widetilde\jmath=0,\ldots,\widetilde p-1$ and $\widetilde k=0,\ldots,\widetilde q-1$,
and vector of integers $m_{\,\widetilde\jmath\:\widetilde k\,}$, such that $\widetilde p\cdot\widetilde q=N$ and
\be
\label{approx}
	\frac{\widetilde\jmath}{\widetilde p}+\frac{\widetilde k}{\widetilde q}\left(\omega+\frac{\widetilde r}{\widetilde p}\right)
		-\,m_{\,\widetilde\jmath\:\widetilde k\,}\omega\:=\:\frac jp+\frac kq\left(ab\omega+\frac{\widehat r}p\right)\,+\,\mathcal O\left(\,\frac 1q\,\right)
		\quad\mod1\,.	
\ee
Then, we will show that these extra $\mathcal O(1/q)$ terms 
can be neglected without affecting the leading order of the integrand $\mathcal Z$:
\begin{align}
\label{simplification}
\nonumber
	&\sum_{\widetilde\jmath_1,\widetilde\jmath_2=0}^{\widetilde p-1}\:\sum_{\widetilde k_1\ne\widetilde k_2=0}^{\widetilde q-1}
		\log\Gamma_e\left(\Delta+\frac{\widetilde\jmath_1-\widetilde\jmath_2}{\widetilde p}+
		\frac{\widetilde k_1-\widetilde k_2}{\widetilde q}\left(\omega+\frac{\widetilde r}{\widetilde p}\:\!\right)
		-\big(m_{\widetilde\jmath_1\widetilde k_1}-m_{\widetilde\jmath_2\widetilde k_2}\big)\,\omega\:;a\omega,b\omega\right)=\\[2mm]
	&\qquad=\sum_{j_1,j_2=0}^{p-1}\:\sum_{k_1\ne k_2=0}^{q-1}
		\log\Gamma_e\Bigg(\Delta+\frac{j_1-j_2}p+\frac{k_1-k_2}q\left(ab\omega+\frac{\widehat r}p\right);a\omega,b\omega\Bigg)+o(q^2)\:.
\end{align}
The quantity in the second line of (\ref{simplification}) is 
the same as (\ref{Phi}). 
Hence, the rest of the computation is identical to the one in section \ref{large N dominant contribution revisited}.

Let us set $\widehat h\equiv\text{gcd}(ab,q,\,\widehat r\,)\,$; if we reparametrize the index $k$ in terms of new indices
$k'=0,\ldots,q/\;\!\widehat h-1\,$ and $\,k''=0,\ldots,\widehat h-1$ such that $k\equiv k'+(q/\;\!\widehat h)\:\!k''$,
we can write the following:
\begin{alignat}{3}
\nonumber
	&\frac jp+\frac kq\left(ab\omega+\frac{\widehat r}p\right)&&=\,
		\frac jp\,+\,\frac k{q\,/\,\widehat h}\,\bigg(\frac{ab}{\widehat h}\,\omega+\frac{\widehat r/\,\widehat h}p\:\!\bigg)=&&\\[2mm]
	& &&=\,\frac jp\,+\,\frac{k'}{q\,/\,\widehat h}\,\bigg(\frac{ab}{\widehat h}\,\omega+\frac{\widehat r/\,\widehat h}p\:\!\bigg)
		+\frac{k''(\widehat r/\,\widehat h)}p\:&&\mod1,\,\omega\\[2mm]\nonumber
	& &&=\,\frac{j\;\!'}p\,+\,\frac{k'}{q\,/\,\widehat h}\,\bigg(\frac{ab}{\widehat h}\,\omega+\frac{\widehat r/\,\widehat h}p\:\!\bigg)&&\mod1,\,\omega\,.
\end{alignat}
In the last step we defined a new index $j\;\!'\,$ as $\,j\;\!'\equiv j\,+\,k''(\widehat r/\,\widehat h)\mod p$.
The dependence on the index $k''$ has dropped completely modulo $1,\,\omega\,$;
considering that BAE solutions cannot repeat values modulo $1,\,\omega\,$, we will have to reintroduce the dependence on $k''$ 
as a part of the $\mathcal O(1/q)$ term.

As a consequence of the definition of $\,\widehat h\,$, we have that $\text{gcd}(ab/\,\widehat h,\,q/\,\widehat h,\,\widehat r/\,\widehat h)=1$.
Therefore 
if we set $h\equiv\text{gcd}(q/\,\widehat h\,,ab)$, $\,\widetilde p\equiv hp\,$ and $\,\widetilde q\equiv N/\,\widetilde p\,$,
we can find indices $\widetilde\jmath=0,\ldots,\widetilde p-1\,$, 
$\,\widehat k=0,\ldots,\widetilde q\,/\,\widehat h-1$ 
and an integer $\,\widetilde r\,$ such that
\be
	\frac{j\;\!'}p\,+\,\frac{k'}{q\,/\,\widehat h}\,\bigg(\frac{ab}{\widehat h}\,\omega+\frac{\widehat r/\,\widehat h}p\:\!\bigg)\:=\:\,
		\frac{\widetilde\jmath}{\widetilde p}\,+\,\frac{\widehat k}{\widetilde q\,/\,\widehat h}\left(\omega+\frac{\widetilde r}{\widetilde p}\right)
		\mod1,\,\omega	\,.
\ee
This relation can be obtained by following the same steps used to prove (\ref{idealscenario}),
with $\,ab/\,\widehat h$, $\,q/\,\widehat h\,$ and $\,\widehat r/\,\widehat h\,$ taking the place of $ab$, $q$ and $r$ respectively.

We can now chose the value for the vector of integers $m$ so that the following identity holds:
\be
\label{approx_m}
	\frac jp+\frac kq\left(ab\omega+\frac{\widehat r}p\right)\:\equiv\:\,
		\frac{\widetilde\jmath}{\widetilde p}\,+\,\frac{\widehat k}{\widetilde q\,/\,\widehat h}\left(\omega+\frac{\widetilde r}{\widetilde p}\right)
		+m_{\widetilde\jmath\,\widehat k\,}\omega\mod1\,.
\ee
Lastly, we define the index $\widetilde k=0,\ldots,\widetilde q-1\,$ as $\,\widetilde k\equiv k''+\,\widehat k\,\widehat h\,$, which gives us
\be
\label{approx_last_step}
	\frac{\widetilde\jmath}{\widetilde p}+\frac{\widetilde k}{\widetilde q}\left(\omega+\frac{\widetilde r}{\widetilde p}\right)
		\:\equiv\:\,\frac{\widetilde\jmath}{\widetilde p}\,+\,\frac{\widehat k}{\widetilde q\,/\,\widehat h}\left(\omega+\frac{\widetilde r}{\widetilde p}\right)
		+\frac{k''}{\widetilde q}\left(\omega+\frac{\widetilde r}{\widetilde p}\right)\,.
\ee
Considering that $k''/\,\widetilde q\,=\,\mathcal O(1/q)$,
if we combine relations (\ref{approx_m}) and (\ref{approx_last_step}) together we finally obtain (\ref{approx}).
The only thing left to do is to verify that the simplification (\ref{simplification}) 
works at leading order.

Let us set $Z\equiv\Delta+(\widetilde\jmath_1-\widetilde\jmath_2)/\,\widetilde p\,$. We need to verify that the following is true:
\begin{align}
\label{simplification2}
\nonumber
	&\sum_{\widehat k_1\ne\widehat k_2=0}^{\widetilde q\,/\,\widehat h\,-1}
		\log\Gamma_e\Bigg(Z+\frac{\widehat k_1-\widehat k_2}{\widetilde q\,/\;\!\widehat h}\left(\omega+\frac{\widetilde r}{\widetilde p}\right)
		+\frac{k''_1-k''_2}{\widetilde q}\left(\omega+\frac{\widetilde r}{\widetilde p}\right)
		-\big(m_{\widetilde\jmath_1\widehat k_1}-m_{\widetilde\jmath_2\widehat k_2}\big)\,\omega\:;a\omega,b\omega\Bigg)=\\[1mm]
	&\qquad=\sum_{\widehat k_1\ne\widehat k_2=0}^{\widetilde q\,/\,\widehat h\,-1}
		\log\Gamma_e\Bigg(Z+\frac{\widehat k_1-\widehat k_2}{\widetilde q\,/\;\!\widehat h}\left(\omega+\frac{\widetilde r}{\widetilde p}\right)
		-\big(m_{\widetilde\jmath_1\widehat k_1}-m_{\widetilde\jmath_2\widehat k_2}\big)\,\omega\:;a\omega,b\omega\Bigg)
		+o(q^2)\,.
\end{align}
We are ignoring the sums over $\,\widetilde\jmath_1,\widetilde\jmath_2=0,\ldots,\widetilde p-1\,$ and $\,k''_1,k''_2=0,\ldots,\widehat h-1\,$
because $\,\widetilde p,\,\widehat h\,\sim\,\mathcal O(1)\,$;
if (\ref{simplification2}) holds, then (\ref{simplification}) would immediately follow.

A relation similar to (\ref{simplification2}), albeit simpler, has already been proven in \cite{Benini:2020gjh}, and we can use it as a staring point.
Let us define the following function:
\be
	f(z;\tau)=\sum_{\gamma\ne\delta=1}^{\widetilde N}\log\Gamma_e\Big(z+\frac{\gamma-\delta}{\widetilde N}\,\tau\,;n\tau,n\tau\Big)\,,
\ee
where $n$ is any positive integer. 
As long as $\,z+t\,\tau$ does not cross a zero or a pole of $\Gamma_e$ for any $t\in(-1,0)\cup(0,1)$,
this function 
has been shown to satisfy the following bound:
\be
\label{old_simplification}
	\Big|f\big(z+C\tau/{\widetilde N}\,;\tau\big)-f(z;\tau)\Big|\leq\mathcal O(\widetilde N\log\widetilde N)\,,
\ee
for any $C\in(-1,1)$.
There are a few details about the proof of (\ref{old_simplification}) that will be useful; let us review them briefly.
The first step in the proof 
is to use the mean value theorem to 
write 
\be
\label{mean_value}
	\Big|f\big(z+C\tau/{\widetilde N}\,;\tau\big)-f(z;\tau)\Big|\,\leq\,\frac{|\tau|}{\widetilde N}\,
		\bigg(\Big|\partial_zf\big(z+\bar c_1\tau/{\widetilde N}\,;\tau\big)\Big|+\Big|\partial_zf\big(z+\bar c_2\tau/{\widetilde N}\,;\tau\big)\Big|\bigg)
\ee
for some $\bar c_1, \bar c_2\in\mathbb R$, with $|\bar c_{1,2}|<|C|\,$.%
\footnote{The mean value theorem is applied to the real and imaginary part separately, which is the reason for the need of two constants,
		$\bar c_1$ and $\bar c_2$.} 
Then 
the authors of \cite{Benini:2020gjh} have shown that for any $\bar c\in(-1,1)$ the following bound holds:
\be
\label{old_simplification2}
	\frac1{\widetilde N}\,\Big|\partial_zf\big(z+\bar c\,\tau/{\widetilde N}\,;\tau\big)\Big|\leq
		\frac1{\widetilde N}\sum_{\gamma\ne\delta=1}^{\widetilde N}
		\left|\frac{\partial_z\Gamma_e\Big(z+\frac{\gamma-\delta+\bar c}{\widetilde N}\,\tau\,;n\tau,n\tau\Big)}
		{\Gamma_e\Big(z+\frac{\gamma-\delta+\bar c}{\widetilde N}\,\tau\,;n\tau,n\tau\Big)}\right|\leq\mathcal O(\widetilde N\log\widetilde N)\,.
\ee
Relation (\ref{old_simplification}) then follows from (\ref{mean_value}) and (\ref{old_simplification2}).

We can't use formula (\ref{old_simplification}) to prove (\ref{simplification2}) directly; we need to generalize (\ref{old_simplification}) a bit first.
Given any two subsets $S,S'$ of the set $\{1,\ldots,\widetilde N\}$, we consider the following function:
\be
	f_{S,S'}(z;\tau)=\underset{\gamma\ne\delta}{\sum_{\gamma\in S,\:\delta\in S'}^{\widetilde N}}
		\log\Gamma_e\Big(z+\frac{\gamma-\delta}{\widetilde N}\,\tau\,;n\tau,n\tau\Big)\,.
\ee
Then a similar relation to (\ref{old_simplification}) holds for $f_{S,S'}$ as well:
\be
\label{generalized_simplification}
	\Big|f_{S,S'}\big(z+C\tau/{\widetilde N}\,;\tau\big)-f_{S,S'}(z;\tau)\Big|\leq\mathcal O(\widetilde N\log\widetilde N)\,.
\ee
Indeed, the following trivial inequality:
\be
	\underset{\gamma\ne\delta}{\sum_{\gamma\in S,\:\delta\in S'}^{\widetilde N}}\left|
		\frac{\partial_z\Gamma_e\Big(z+\frac{\gamma-\delta+\bar c}{\widetilde N}\,\tau\,;n\tau,n\tau\Big)}
		{\Gamma_e\Big(z+\frac{\gamma-\delta+\bar c}{\widetilde N}\,\tau\,;n\tau,n\tau\Big)}\right|\leq
		\sum_{\gamma\ne\delta=1}^{\widetilde N}\left|\frac{\partial_z\Gamma_e\Big(z+\frac{\gamma-\delta+\bar c}{\widetilde N}\,\tau\,;n\tau,n\tau\Big)}
		{\Gamma_e\Big(z+\frac{\gamma-\delta+\bar c}{\widetilde N}\,\tau\,;n\tau,n\tau\Big)}\right|
\ee
together with (\ref{old_simplification2}) and 
an analogue of (\ref{mean_value}) imply (\ref{generalized_simplification}).

We can use relation (\ref{generalized_simplification}) to show that
\bea 
\label{generalized_simplification2}
	\,&\sum_{\gamma\ne\delta=1}^{\widetilde N}
		\log\Gamma_e\Big(z+\frac{\gamma-\delta}{\widetilde N}\,\tau-(m_\delta-m_\gamma)\,\omega+C\tau/{\widetilde N}\,;n\tau,n\tau\Big)=\\
	\,&\qquad=\sum_{\gamma\ne\delta=1}^{\widetilde N}
		\log\Gamma_e\Big(z+\frac{\gamma-\delta}{\widetilde N}\,\tau-(m_\delta-m_\gamma)\,\omega\,;n\tau,n\tau\Big)
		+\mathcal O(\widetilde N\log\widetilde N)\,,
\eea 
where 
$\{m_\delta\}_{\delta=1}^{\widetilde N}$ is a vector of integers between 1 and $ab$ (with $\,ab\sim\mathcal O(\widetilde N^0)\,$),  $\omega\in\mathbb C$
and $z$ is such that $z-(m_\delta-m_\gamma)\,\omega+t\,\tau$ does not cross a zero or a pole of $\Gamma_e$ for any $t\in(-1,0)\cup(0,1)$
and any possible value of $(m_\delta-m_\gamma)$. Indeed, we can write
\be
	\sum_{\gamma\ne\delta=1}^{\widetilde N}
		\log\Gamma_e\Big(z+\frac{\gamma-\delta}{\widetilde N}\,\tau-(m_\delta-m_\gamma)\,\omega\,;n\tau,n\tau\Big)\equiv
	\sum_{i_1,i_2=1}^{ab}f_{S(i_1),S(i_2)}\big(z-(i_1-i_2)\,\omega\,;\tau\big)\,,
\ee
where $S(i)\equiv\{\delta\:|\:m_\delta=i\}$. Then (\ref{generalized_simplification2}) follows directly from (\ref{generalized_simplification}).

At last, let us show the validity of simplification (\ref{simplification2}). 
In order to apply (\ref{generalized_simplification2}) we need to first 
change the moduli of the $\Gamma_e$ from $(a\omega,b\omega)$ to
$\big(n(\omega+\widetilde r\,/\,\widetilde p),\,n(\omega+\widetilde r\,/\,\widetilde p)\big)$, 
where $n$ is some positive integer.
We can use the following identity:
\be
	\log\Gamma_e\Big(u\,;a\omega,b\omega\Big)=
		\sum_{\ell_1=0}^{b\:\!\widetilde p-1}\,\sum_{\ell_2=0}^{a\:\!\widetilde p-1}
		\log\Gamma_e\Big(u+(\ell_1a+\ell_2b)\:\!\omega\,;\;\!\widetilde p\:\!ab\omega+abr,\;\!\widetilde p\:\!ab\omega+abr\Big)\,,
\ee
which follows from (\ref{Gamma_prod1}) and the invariance of $\Gamma_e$ under integers shifts.
Then, for any given value of $\ell_1,\ell_2$ we can use (\ref{generalized_simplification2}) with
$z\equiv Z+(\ell_1a+\ell_2b)\,\omega\,$,
$\tau\equiv\omega+\widetilde r\,/\,\widetilde p\,$ and $\,n\equiv\widetilde p\,ab$.
It is easy to verify that $z$ satisfies the required condition necessary for avoiding zeros and poles,
considering that the possible values for $\Delta$ are $\Delta\equiv\tau+\sigma$ and $\Delta\equiv\Delta_I$, with
$\Delta_I$ satisfying condition (\ref{Stokes}).
This concludes the proof of (\ref{simplification2}).

\bibliographystyle{ytphys}
\bibliography{BHentropy}

\end{document}